\documentclass[aps,prd,twocolumn,preprintnumbers,superscriptaddress,nofootinbib,floatfix]{revtex4-1}

\usepackage{hyperref}

\usepackage[dvipsnames]{xcolor}
\usepackage{amsmath}
\usepackage{bm}
\usepackage{comment}
\usepackage{amsfonts}
\usepackage{graphicx}
\usepackage{epstopdf}
\usepackage{hyperref}
\usepackage{array}
\usepackage[utf8]{inputenc}
\usepackage{soul}
\usepackage{color}
\usepackage[T1]{fontenc}
\definecolor{steelblue}{RGB}{25,25,112}
\definecolor{dullblue}{rgb}{0,0.298,0.49}
\definecolor{darkred}{rgb}{0.545,0,0}
\definecolor{blue2}{cmyk}{1, 0.1, 0.1, 0}

\hypersetup{colorlinks,linkcolor={darkred},citecolor={dullblue},urlcolor={dullblue}}  

\enlargethispage{\baselineskip}

\newcommand{\MC}{M_\mathrm{AMC}}
\newcommand{\RC}{R}
\newcommand{\rC}{\rho}

\newcommand{\be}{\begin{equation}}
\newcommand{\ee}{\end{equation}}
\newcommand{\bea}{\begin{eqnarray}}
\newcommand{\eea}{\end{eqnarray}}

\usepackage{colortbl}
\definecolor{blue2}{cmyk}{1, 0.1, 0.1, 0}

\begin{document}

\title{Transient Radio Signatures from Neutron Star Encounters \\ with QCD Axion Miniclusters}

\newcommand{\GRAPPA}{\affiliation{Gravitation Astroparticle Physics Amsterdam (GRAPPA), Institute for Theoretical Physics Amsterdam and Delta Institute for Theoretical Physics, University of Amsterdam, Science Park 904, 1098 XH Amsterdam, The Netherlands}}
\newcommand{\OKC}{\affiliation{The Oskar Klein Centre for Cosmoparticle Physics,
    AlbaNova University Center,\\
	Roslagstullsbacken 21,
	SE--106\.91 Stockholm,
	Sweden}} 
\newcommand{\IFCA}{\affiliation{Instituto de F\'isica de Cantabria (IFCA, UC-CSIC), Av.~de
Los Castros s/n, 39005 Santander, Spain}}
\newcommand{\INFN}{\affiliation{INFN, Laboratori Nazionali di Frascati, C.P. 13, 100044 Frascati, Italy}}
\author{Thomas D. P. Edwards}\email[Electronic address: ]{thomas.edwards@fysik.su.se} \OKC \GRAPPA
\author{Bradley J. Kavanagh}\email[Electronic address: ]{kavanagh@ifca.unican.es} \IFCA \GRAPPA
\author{Luca Visinelli}\email[Electronic address: ]{luca.visinelli@sjtu.edu.cn} \GRAPPA \INFN
\author{Christoph Weniger}\email[Electronic address: ]{c.weniger@uva.nl} \GRAPPA

\date{\today}

\begin{abstract}
The QCD axion is expected to form dense structures known as axion miniclusters if the Peccei-Quinn symmetry is broken after inflation. Miniclusters that have survived until today will interact with neutron stars (NSs) in the Milky Way to produce transient radio signals from axion-photon conversion in the NS magnetosphere. We quantify the properties of these encounters and find that they occur frequently ($\mathcal{O}(1-100)\,\mathrm{day}^{-1}$); last between a day and a few months; are spatially clustered towards the Galactic center; and can reach observable fluxes. These radio transients are within reach of current generation telescopes and therefore offer a promising pathway to discovering QCD axion dark matter.
\end{abstract}

\maketitle

\textbf{\textit{Introduction ---}} Peccei-Quinn (PQ) theory~\cite{Peccei:1977hh, Peccei:1977ur} predicts the existence of the QCD axion~\cite{Weinberg:1977ma, Wilczek:1977pj}, which could simultaneously solve the strong-CP problem and act as a compelling candidate for particle dark matter (DM)~\cite{Vilenkin:1981kz, Abbott:1982af, Dine:1982ah, Preskill:1982cy}.
The QCD axion is the pseudo-Nambu-Goldstone boson~\cite{Nambu:1960tm,Goldstone:1961eq,Goldstone:1962es} of the new global PQ symmetry. 
Laboratory searches are underway worldwide to directly detect this QCD axion~\cite{ Sikivie:1983ip, Sikivie:1985yu,Irastorza:2018dyq, Sikivie:2020zpn}. Astrophysical observations are also a promising avenue for detecting the axion~\cite{Cadamuro:2012rm,Ayala:2014pea,2015JCAP...10..015V,Lee:2018lcj,Calore:2020tjw}. In particular, radio observations could be used to look for emission from the conversion of axions into photons in neutron star (NS) magnetospheres~\cite{Pshirkov:2007st, Huang:2018lxq, Hook:2018iia, Safdi:2018oeu, Edwards:2019tzf,Foster:2020pgt,Darling:2020uyo}. In this Letter, we propose and characterize a new class of radio source, arising from encounters between NSs and overdense structures known as axion {\it miniclusters} (AMCs)~\cite{Hogan:1988mp, Kolb:1993zz, Kolb:1993hw, Kolb:1994fi, Kolb:1995bu, Eggemeier:2019khm}.\footnote{We use the terms `miniclusters' and AMCs interchangeably.}

As axions fall towards an NS, they can resonantly convert into photons within the magnetosphere. This occurs at a radius $R_c$ at which the plasma frequency $\omega_p$ in the magnetosphere equals the axion mass $m_a$~\cite{Hook:2018iia}. Assuming a Goldreich-Julian model for the NS magnetosphere~\cite{Goldreich:1969sb}, the power radiated per unit solid angle is derived in the WKB and stationary phase approximations as~\cite{Pshirkov:2007st, Huang:2018lxq, Hook:2018iia, Safdi:2018oeu}
\begin{equation}
\label{eq:power_radiated}
\frac{\mathrm{d}\mathcal{P}_a}{\mathrm{d}\Omega} \sim \frac{\pi}{3}\,g_{a\gamma\gamma}^2B_0^2\,\frac{R_{\rm NS}{}^6}{R_c{}^3}\,\frac{\rho_a}{m_a}\,,
\end{equation}
where $\rho_a$ is the axion density at the conversion radius~\cite{Hook:2018iia}, $g_{a\gamma\gamma}$ is the axion-photon coupling, and we have averaged over viewing angles. We also fix the rotation axis to be aligned with the NS dipole field and set the NS radius $R_{\rm NS} = 10\,$km (see Supplemental Material for details of the NS modeling). The power scales with $B_0^2$ --- where $B_0$ is the magnetic field strength at the NS poles --- emphasizing why NSs are the most promising astrophysical target for these searches, with the highest known magnetic fields in the Universe $\mathcal{O}(10^{10}-10^{15})$~G~\cite{Phinney:1994gf}. The flux also scales with $\rho_a$, meaning that regions of large axion density --- such as AMCs --- can give rise to very bright radio sources.

AMCs are a generic feature of models in which the PQ symmetry is broken after the end of inflation~\cite{Kolb:1995bu}. Large spatial variations of the axion density around the QCD epoch
lead to the formation of minicluster `seeds'~\cite{Vaquero:2018tib} which collapse into gravitationally bound AMCs around matter-radiation equality~\cite{Zurek:2006sy}. This evolution has been confirmed by numerical simulations, which show that a significant fraction of DM axions might be contained within such bound structures~\cite{Vaquero:2018tib, Buschmann:2019icd}.
Despite their low mass ($10^{-19}\,M_{\odot} \lesssim \MC \lesssim 10^{-5}\,M_{\odot}$) and large radius ($10^{-8}\,\mathrm{pc}\lesssim\RC \lesssim 10^{-2}\,\mathrm{pc}$), the density of an AMC can be many orders of magnitude larger than the local DM density~\cite{Visinelli:2018wza}.
On the other hand, AMCs are significantly more diffuse than stars. In our companion paper~\cite{Kavanagh:2020gcy} (hereafter KEVW20), we show that tidal interactions with stars can have a dramatic effect on the survival of AMCs in the Milky Way (MW). AMCs towards the inner regions of the Galaxy undergo significant stripping and disruption, whereas those further out remain intact. This in turn is reflected in the observational signatures of AMCs.

Here, we build upon the results of KEVW20 to predict the rate, brightness, and  sky distributions of encounters between AMCs and NSs. For a typical MW virial velocity of $v \sim 200 \,\mathrm{km/s} \sim 10^{-11}\,\mathrm{pc/s}$, the time taken for a NS to pass through an AMC is expected to be $\mathcal{O}(10^3-10^9)\,$s, meaning that these interactions would appear as radio \textit{transients}. As we will show, this is particularly true for the brightest events which last between $10^5$ to $10^7$ seconds. 
We consider a Kim-Shifman-Vainshtein-Zakharov (KSVZ)-like QCD axion~\cite{Kim:1979if, Shifman:1979if} of mass $m_a = 20\,\mu$eV, motivated by recent simulations~\cite{Klaer:2017ond, Buschmann:2019icd}. This corresponds to a radio frequency of $f = 4.84 \,\mathrm{GHz}$ and an axion-photon coupling of $g_{a\gamma\gamma} \approx 8\times10^{-15}{\rm\,GeV}^{-1}$. All code associated with this work is available online at \href{https://github.com/bradkav/axion-miniclusters/}{github.com/bradkav/axion-miniclusters}~\cite{AMC_code}.

\vskip 3pt

{\bf \textit{Axion Miniclusters in the Milky Way}\;---}
Due to the randomness of the initial overdensity fluctuations, miniclusters are born with a wide range of masses and densities.
During matter-domination, the minicluster halo mass function (HMF) evolves under hierarchical structure formation, allowing ever heavier AMCs to form.
$N$-body simulations modeling AMC evolution 
from recombination to $z\approx 99$ predict a featureless HMF with a characteristic slope $\mathrm{d}n/\mathrm{d}\log M \sim M^{-0.7}$~\cite{Eggemeier:2019khm}, corroborating semi-analytic studies~\cite{Fairbairn:2017sil, Fairbairn:2017dmf, Ellis:2020gtq}. We consider AMC masses between $3.3 \times 10^{-19}\leq \MC/M_\odot \leq  5.1 \times 10^{-5}$~\cite{Fairbairn:2017sil, Fairbairn:2017dmf} and assume that these AMCs make up 100\% of the DM, tracing the NFW profile of the MW halo.
Miniclusters are also characterized by their overdensity parameter $\delta$, which depends upon the random initial conditions of the axion field and its gradient at the onset of axion oscillations~\cite{Kolb:1993zz, Kolb:1993hw, Kolb:1994fi}. We take the distribution of $\delta$ from recent simulations~\cite{Buschmann:2019icd} and map it to the AMC characteristic density $\rho_\mathrm{AMC}$ as in Ref.~\cite{Kolb:1994fi}.

Tidal interactions between AMCs and their local environment can have a significant effect on AMC properties~\cite{Zhao:2005py, Tinyakov:2015cgg, Berezinsky:2013fxa,2017JETP..125..434D, Kavanagh:2020gcy}.
Stellar encounters prove to be the most important and can easily lead to the total disruption of AMCs. In addition, many successive weak encounters can cause surviving AMCs to lose mass, as well as altering their internal density.
In KEVW20, we present Monte Carlo simulations used to assess the effects of these stellar tidal interactions, starting from the initial distribution of masses and densities described above. These simulations allow us to describe the properties of AMCs across the MW today.

The internal density profiles of AMCs are not well understood. For example, Ref.~\cite{Eggemeier:2019khm} finds that AMCs with masses $\MC \gtrsim 10^{-13}\,M_\odot$ have approximately Navarro-Frenk-White (NFW)~\cite{Navarro:1995iw} profiles in their outer regions whereas lighter AMCs are expected to have Power-law (PL) profiles~\cite{Zurek:2006sy}. 
For each simulation we therefore treat the entire population of AMCs as having a single universal structure given by either an NFW or a PL density profile (see Supplemental Material for examples). For a fixed characteristic density $\rho_\mathrm{AMC}$, the \textit{mean} internal density $\bar{\rho}$ of our assumed NFW profile is $\mathcal{O}(10^5)$ times lower than for the PL profile. This leads to quantitative differences in the survival probability and distributions of masses and radii. Using these two density profiles therefore allows us to generously bound the uncertainties coming from the internal AMC structure. 

Another source of uncertainty is related to the formation of axion stars (ASs). These are non-relativistic compact objects, described by solutions to the Schr\"{o}dinger-Poisson equation~\cite{Kolb:1993zz, Seidel:1993zk}, which can potentially form in the centers of miniclusters~\cite{Levkov:2018kau, Eggemeier:2019jsu, Chen:2020cef}. ASs have an inverse relationship between their mass and radius, leading to a potentially problematic scenario for a low-mass AMC in which its radius is \textit{smaller} than that of the AS in its center. To avoid this issue, we first follow the evolution of all AMCs, described initially by the HMF above, then apply a cut to remove these potentially problematic light AMCs.\footnote{In KEVW20, we also include an initial mass-loss of $5-40\%$ for AMCs with NFW profiles, due to tidal stripping from the DM halo of the MW.} 
The remaining miniclusters form our fiducial sample and are used throughout the rest of this work (see Supplemental Material and KEVW20 for further details).

\vskip 3pt
{\bf \textit{Minicluster - Neutron Star Encounters}\;---} 
The AMC-NS encounter cross section is given by $\sigma(u) = \pi \RC^2\left(1 + 2GM_{\rm NS}/(\RC u^2)\right)$, where $R$ is the minicluster radius and $M_{\rm NS} = 1.4\,M_\odot$ is the NS mass.
This expression includes a gravitational focusing term~\cite[p.~627]{BinneyTremaine:2008} that depends on the relative velocity of the encounter $u$.
We assume that the velocity dispersion of the NS and minicluster populations is $\sigma_v(r) = \sqrt{G M_\mathrm{encl}(r)/r} \approx 200 \,\mathrm{km/s}$, depending on the enclosed MW mass $M_\mathrm{encl}(r)$ within a galactocentric radius $r$. The encounter velocity then follows a Maxwell-Boltzmann distribution $f_u(u)$ with dispersion $\sigma_u(r) = \sqrt{2}\sigma_v(r)$~\cite[Problem 8.8]{BinneyTremaine:2008}. The velocity-weighted cross section can then be written:
\begin{equation}
\label{eq:cross_section_u}
\langle\sigma u \rangle(r) \!\equiv\! \int \mathrm{d}^3u f_u(u)u\sigma(u) \approx \sqrt{8\pi}\sigma_u(r) \RC^2\!\left(1\!+\!\hat R/\RC\right),
\end{equation}
where $\hat R = GM_{\rm NS}/\sigma_u^2 \sim \mathcal{O}(10^{-7}\,\mathrm{pc})$.

\begin{figure}[t!]
\begin{center}
\includegraphics[width=0.49\textwidth]{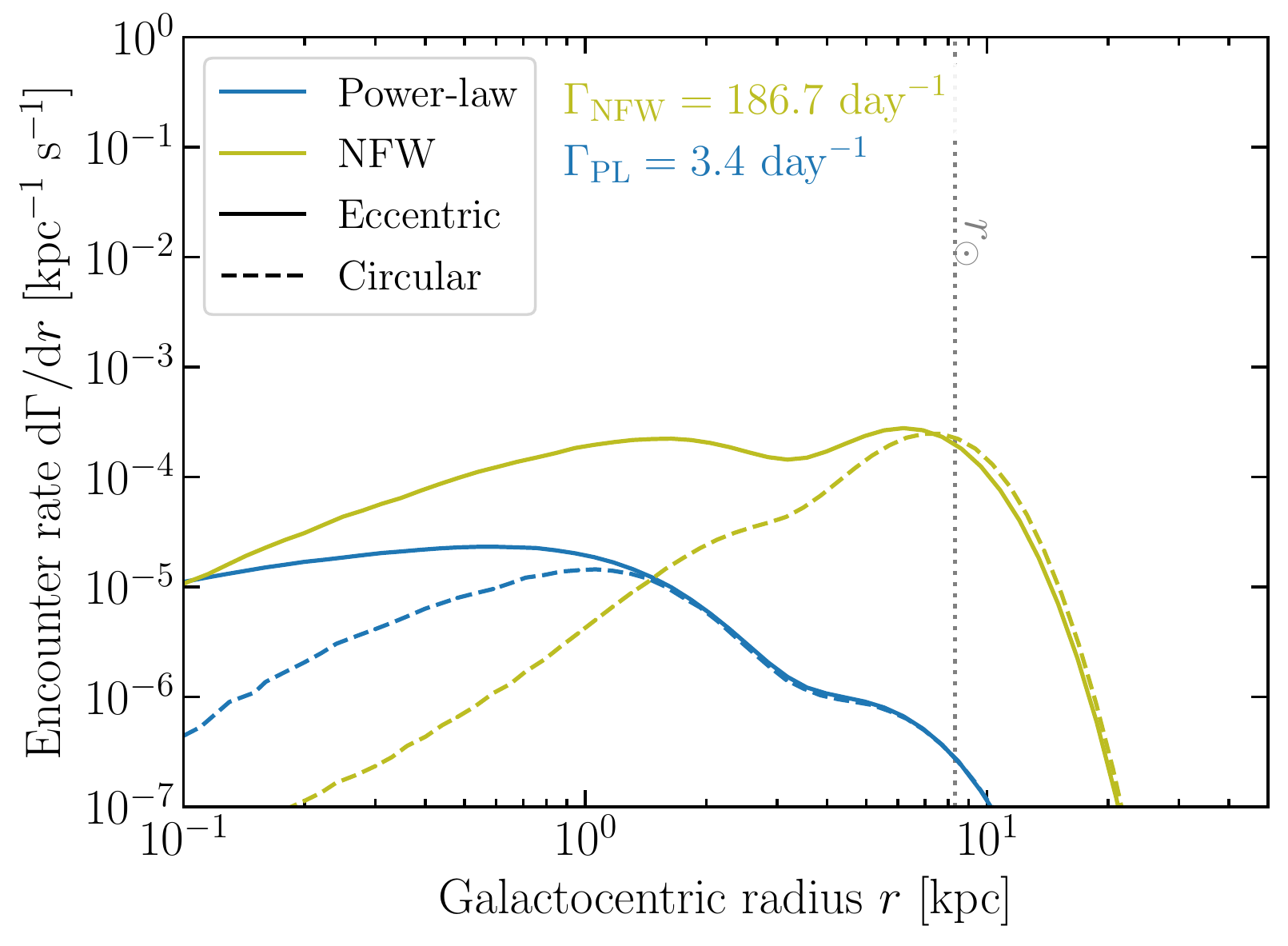}
\caption{AMC-NS encounter rate incorporating the effects of AMC disruptions due to stellar encounters. We show results assuming that AMCs are solely on circular orbits (dashed) and assuming a distribution of orbital eccentricities (dotted). The vertical dotted line marks the position of the Solar System, $r_\odot$.}
\label{fig:PDF}
\end{center}
\end{figure}

Interactions on the outskirts of large, diffuse AMCs dominate the encounter rate but do not produce a significant increase in the axion density close to the NS. 
We therefore consider only interactions with impact parameters $b < b_\mathrm{cut}$ (with $b_\mathrm{cut} \leq R$), such that the peak overdensity during the encounter is at least 10\% of the local DM density $\rho_\mathrm{DM}(r)$. We parametrize this in terms of an \textit{effective} cross section $\langle\widetilde{\sigma u} \rangle(r)$ which
saturates at the standard cross section $\langle\sigma u \rangle(r)$ for dense AMCs. The expected rate of encounters over the entire MW is then~\cite{Buckley:2020fmh,Prabhu:2020yif}

\begin{equation}
\label{eq:rate}
\Gamma = \int \mathrm{d}^3{\bf r} \int \mathrm{d}\RC\, \frac{\mathrm{d}n_\mathrm{AMC}(r)}{\mathrm{d}\RC}\, n_{\rm NS}({\bf r})\,\langle\widetilde{\sigma u} \rangle(r) \,,
\end{equation}
where $n_{\rm NS}({\bf r})$ is the NS number density at position ${\bf r}$ and $\mathrm{d}n_\mathrm{AMC}(r)/\mathrm{d}\RC$ is the differential number density of AMCs with radius $R$, computed in KEVW20. 
Taking into account the distribution of AMC properties, we find that $\mathrm{d}\Gamma/\mathrm{d}M_\mathrm{AMC} \sim 1/M_\mathrm{AMC}$, rising more steeply than this at low AMC masses, where the gravitational focusing effect becomes important for $R \lesssim \hat R$.
We assume a population of  $10^9$ NSs in the MW, with 60\% formed in the bulge and 40\% in the disk~\cite{Ofek:2009wt, Sartore:2009wn}, of which 20\% have become unbound due to natal kicks~\cite{Sartore:2009wn}. Explicit expressions for $n_\mathrm{NS}(\mathbf{r})$ are given in the Supplemental Material.

In Fig.~\ref{fig:PDF}, we show the integrand $\mathrm{d}\Gamma/\mathrm{d}r$ of Eq.~\eqref{eq:rate}.
 At the largest radii ($r \gtrsim 10-20\,\mathrm{kpc}$), encounters are rare due to the falling number of both AMCs and NSs. Near the Galactic center, the densities of both AMCs and NSs instead rise rapidly. However, the encounter rate is suppressed by the low survival probability of AMCs in this dense environment, leading to a plateau.\footnote{Note that the dip at 3-4\,kpc is a coincidence between the falling survival probability towards the Galactic center and scale at which the bulge population of NSs becomes dominant.} We find that the encounter rate is larger for eccentric than for circular orbits; AMCs on eccentric orbits spend less time at small radii, leading to a larger survival probability.
For NFW profiles, there is a comparable contribution from encounters with bulge NSs at small radii and encounters with disk NSs at larger radii (close to the Solar circle $r_\odot$). Miniclusters with PL profiles are more dense and therefore smaller than those with NFW profiles, leading to an overall decrease in the encounter rate. However, these dense AMCs are also more resistant to disruption in the Galactic center, leading to a greater survival probability at small $r$. This compensates for their smaller size and means that for PL miniclusters most encounters occur with NSs in the bulge.
Over the entire Galaxy, we expect encounter rates of $\Gamma_\mathrm{PL} = 3.4$~day$^{-1}$ and $\Gamma_\mathrm{NFW} = 186.7$~day$^{-1}$. If we had neglected the stellar disruption of AMCs described in KEVW20, these encounter rates would be larger by a factor of 1.4 and 45.4 for PL and NFW profiles respectively.

\begin{figure*}[t!]
\includegraphics[width=0.99\textwidth]{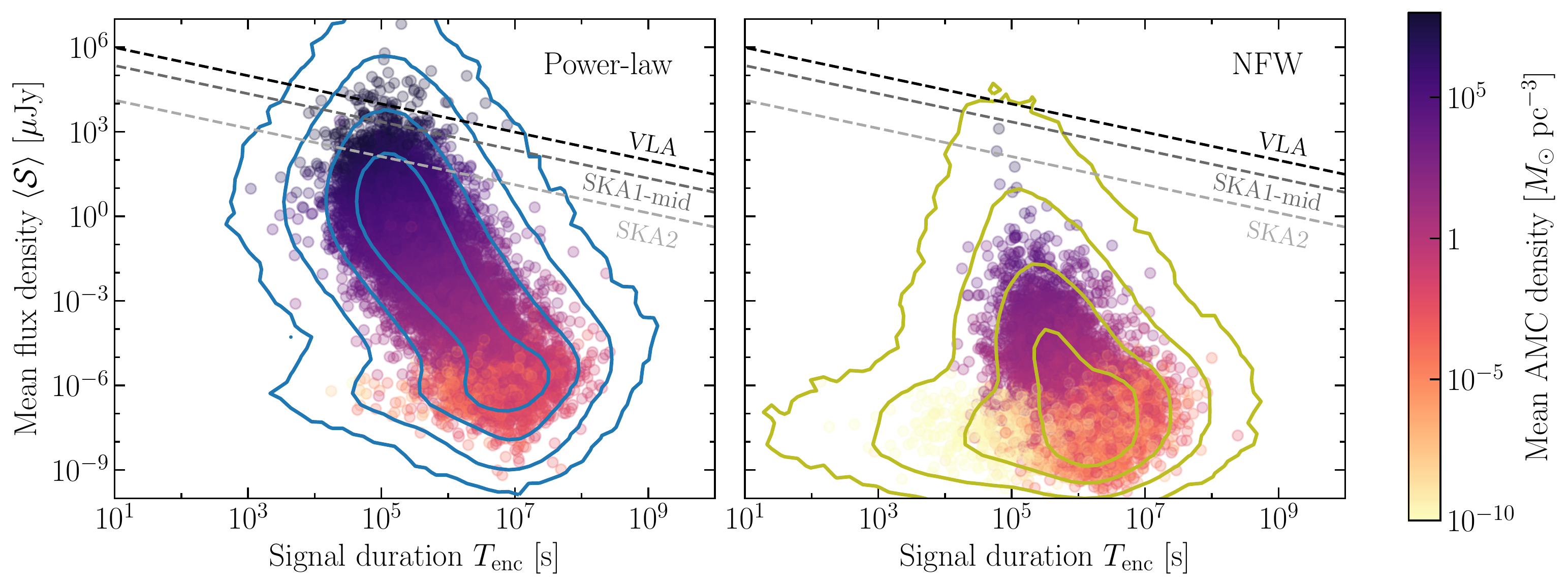}
\caption{Mean flux density and duration of the radio signal for a random sample of $10^4$ AMC-NS encounters, colored by mean AMC density. Overlaid are 1-, 2-, 3- and 4-$\sigma$ contours for the distribution of $\langle \mathcal{S} \rangle$ and $T_\mathrm{enc}$, derived from a larger sample of $10^7$ simulations.
The diagonal dashed lines show the minimum $5\sigma$ detectable flux estimated using the radiometer equation~\cite{1984bens.work..234D} (where we assume that the observation time equals the duration of the encounter) for the Very Large Array (VLA)~\cite{2011ApJ...739L...1P}, SKA1-mid~\cite{Calore:2015bsx,Gaggero:2016dpq}, and SKA2~\cite{Safdi:2018oeu}. We assume AMCs have Power-law (left) or NFW (right) internal density profiles.}
\label{fig:Flux_Tenc_den}
\end{figure*}

\vskip 3pt

{\bf \textit{Signal Estimation}\;---}
For each choice of minicluster profile, we sample $10^7$ encounters to calculate the expected distributions of fluxes, durations, and sky locations.
We sample the galactocentric radius of the encounter according to the encounter rate $\mathrm{d}\Gamma/\mathrm{d}r$ in Fig.~\ref{fig:PDF}. We draw the height of the encounter $z_\mathrm{cyl}$ above the Galactic plane from the distribution of NSs along the $z_\mathrm{cyl}$-axis (assuming that the AMC distribution is spherically symmetric), and we draw the galactocentric azimuth angle uniformly between 0 and $2\pi$.
We sample the AMC radius $R$ following $\mathrm{d}\Gamma/\mathrm{d}R$ at fixed galactocentric radius and sample the AMC density, given $R$, from the distributions derived in KEVW20.\
The impact parameter $b \in [0, b_\mathrm{cut}]$ is sampled according to $P(b) \propto b$.

The NS magnetic field at the poles $B_0$ and the period $P$ are drawn from log-normal distributions, with mean and dispersion given by $\log_{10}(B/{\rm G}) = 12.65$, $\sigma_B = 0.55$~\cite{FaucherGiguere:2005ny, Bates:2013uma} and  $\log_{10}(P/{\rm ms}) = 2.7$, $\sigma_P = -0.34$~\cite{Lorimer:2006qs} respectively.

Considering the trajectories of individual axions close to the NS, the maximum impact parameter which still crosses the conversion radius $R_c$ is:
\begin{equation}
b_{\max }=R_{c}\sqrt{1+\frac{2 G M_\mathrm{NS}}{u^{2} R_{c}}} \sim \mathcal{O}(10^{-9})\,\mathrm{pc}\,.
\end{equation}
AMCs have radii several orders of magnitude larger than this, so we can consider the NS as tracing the internal AMC density $\rho_\mathrm{int}(R)$ during the encounter. Such a direct encounter is likely to completely disrupt the AMC. However, the relaxation time for the AMC is much longer than the encounter time and we therefore neglect the evolution of the minicluster during the NS transit.

For each encounter, we estimate the radio flux density:
\begin{equation}
\label{eq:signal}
\mathcal{S} = \frac{1}{{\rm BW}}\frac{1}{4\pi s^2}\,\frac{\mathrm{d}\mathcal{P}_a}{\mathrm{d}\Omega}\,,
\end{equation}
where $s$ is the distance of the encounter from Earth. The signal bandwidth BW is typically set by the axion velocity dispersion far from the NS~\cite{Hook:2018iia}, leading to narrow-band line emission. However, because of the small internal velocity dispersion of the AMCs ($\lesssim 1\,\mathrm{km/s}$), this is unlikely to be the main source of the signal bandwidth. We therefore fix the bandwidth of the signal to a larger value, $1\,\mathrm{kHz}$, representative of the resolution of current and planned radio telescopes~\cite{2011ApJ...739L...1P,SKA-design,2019arXiv191212699B}. 
Determining the full directional dependence of the radio emission is highly non-trivial, though there have been a number of recent developments dealing, for example, with non-radial infall of axions~\cite{Leroy:2019ghm, Foster:2020pgt}. Here, we assume for simplicity that the emission is isotropic; Eq.~\eqref{eq:power_radiated} has been averaged over viewing angle and we have fixed $R_c$ to its angular average. If future studies show that the radio emission is instead concentrated in a fraction $f_\mathrm{beam}$ of the unit sphere, then our results can be straightforwardly re-interpreted: the observed rate will be reduced by a factor $\sim f_\mathrm{beam}$ and the flux density increased by a corresponding factor $\sim f_\mathrm{beam}$.

We estimate the mean flux density $\langle \mathcal{S} \rangle$ of each encounter by averaging Eq.~\eqref{eq:signal} over the duration of the encounter $T_\mathrm{enc} = 2\sqrt{R^2 - b^2}/u$. Note that the peak flux during the encounter is generally comparable to the mean flux. 
Since each encounter is independent, the number of expected encounters starting within a time step $\Delta t$ is Poisson distributed with mean $\Gamma\Delta t$. Our simulations can therefore be combined into a time series for the predicted signal by taking a uniform distribution of start times. 
In Fig.~\ref{fig:Flux_Tenc_den}, we show the distributions of $T_\mathrm{enc}$ and $\langle\mathcal{S}\rangle$ derived from our full sample of AMC-NS encounters. We also plot a smaller sample of $10^4$ individual encounters, colored by the mean internal density of the AMC.

For PL AMCs, the distribution of encounters peaks at flux densities between $10^{-6} \, \mu$Jy and $10^2 \, \mu$Jy, with a typical duration of 1-100 days. The distribution also includes a number of bright events $\mathcal{O}(1 \,\mathrm{Jy})$ which should be detectable by current radio telescopes such as the Very Large Array (VLA)~\cite{1983IEEEP..71.1295N}. These brightest events come from encounters with dense AMCs ($\gtrsim 10^5\,M_\odot\,\mathrm{pc}^{-3}$). AMCs with NFW internal density profiles have a density around $10^5$ times smaller than their PL counterparts, making high-flux events rarer.
However, this is partially compensated by the larger encounter rate between NSs and NFW AMCs. 
The rate of encounters above a flux of 1 mJy (a sensitivity which has been achieved in recent searches for NS radio emission~\cite{Foster:2020pgt, Darling:2020uyo, Darling:2020plz}) is $\Gamma_\mathrm{PL}(\Psi > 1\,\mathrm{mJy}) = 0.04/\mathrm{day}$ and $\Gamma_\mathrm{NFW}(\Psi > 1\,\mathrm{mJy}) = 0.007/\mathrm{day}$ for PL and NFW miniclusters respectively. Given the rate and duration of the encounters, we expect at least one bright event in the sky at all times. We also note that the rate of the brightest events ($\langle \mathcal{S}\rangle > 1\,\mathrm{Jy}$) is relatively insensitive to our assumptions on the AMC density profile, once stellar perturbations are taken into account (see Supplemental Material).

\begin{figure}[t!]
\includegraphics[width=\linewidth]{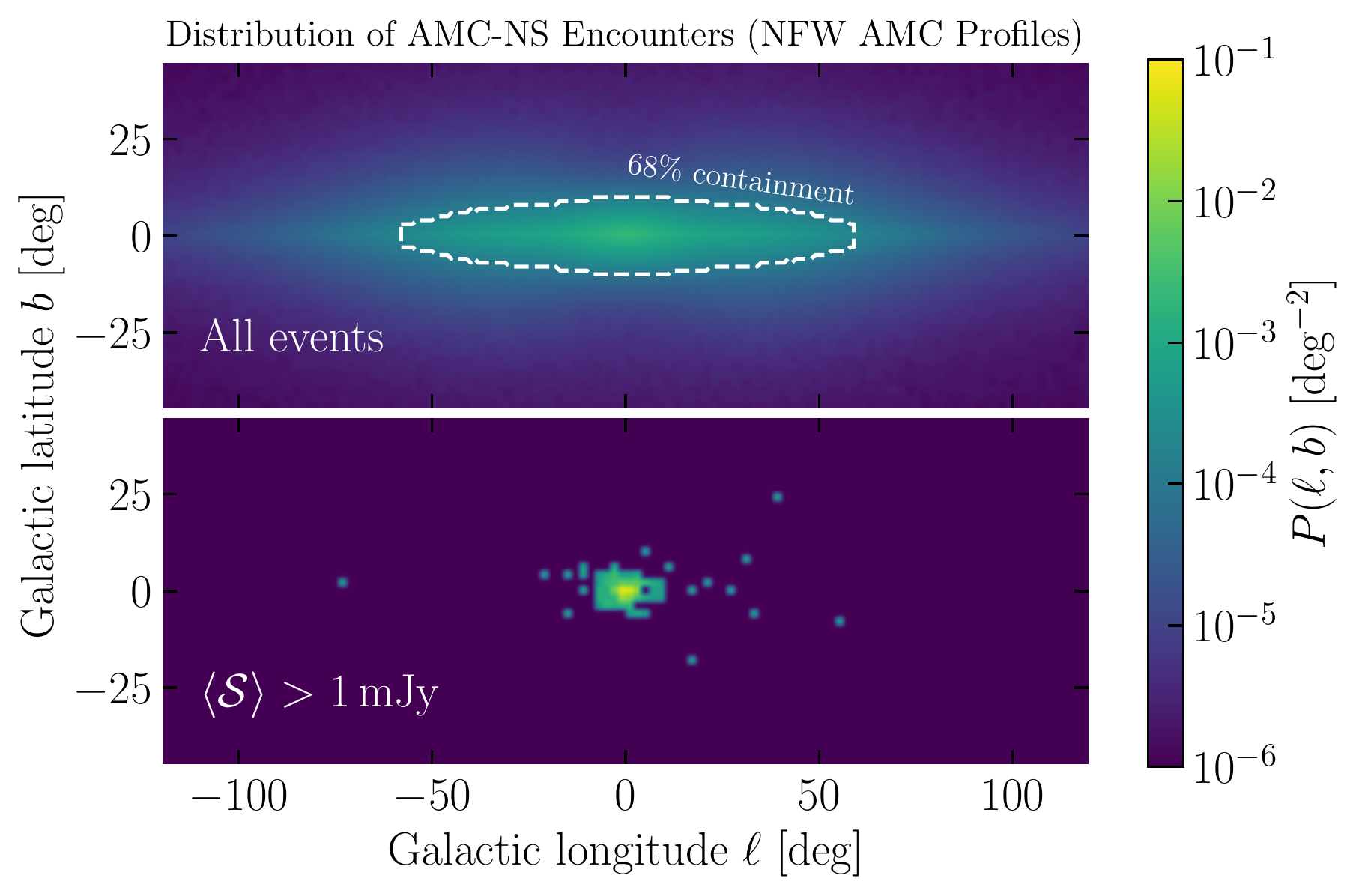}
\caption{Expected distribution of all AMC-NS encounters  (top panel) and bright AMC-NS encounters (bottom panel), assuming NFW internal AMC density profiles. 
For AMCs with PL density profiles, 68\% of all encounters lie within $7^\circ$ of the Galactic center.}
\label{fig:Skymap_moll}
\end{figure}
The sky distribution of AMC-NS encounters is shown in Fig.~\ref{fig:Skymap_moll} for AMCs with NFW profiles. In this case, the encounters occur predominantly towards the Galactic center although there is also a population of events extending along the disk, to a longitude of $|\ell| \lesssim 60^\circ$. This morphology reflects the two populations of NSs in the bulge and disk. In the case of PL AMCs (not shown), encounters are concentrated almost exclusively towards the Galactic center, with 68\% of events lying within $7^\circ$ of the center. Considering only the brightest events, we find that the distributions become even more concentrated towards the Galactic center, as shown in the bottom panel of Fig.~\ref{fig:Skymap_moll}. 

\vskip 10pt

{\bf\textit{Discussion and Conclusion}\;---} In this Letter, we have characterized the radio signatures of axion-photon conversion from encounters between NSs in the Milky Way (MW) and a population of QCD axion miniclusters (see KEVW20~\cite{Kavanagh:2020gcy}). These signatures will appear as regular transient radio point sources (Fig.~\ref{fig:PDF}) with timescales varying from days to over a year.
Interestingly, these transients will be spatially clustered towards the Galactic center (Fig.~\ref{fig:Skymap_moll}) with potentially observable fluxes (Fig.~\ref{fig:Flux_Tenc_den}). This suggests that radio observations could be used to discover QCD axion DM in the near future.

Within the MW, there are a variety of sources of transient radio emission, especially from the Galactic center. A recent analysis of archival VLA data~\cite{Chiti:2016xae} found a number of potential transients within $\sim10^{-3}$ degrees of Sgr A$^{\ast}$ (see also Ref.~\cite{Zhao:2020evb}). Potential explanations of these radio transients are pulsar emission, radio flares from dwarf stars, and outflows from X-ray binaries, all of which emit a broad energy spectrum. AMC-NS encounters could contribute to a population of transients towards the Galactic center. However, our results predict a characteristic line-like emission which would need to be confirmed with dedicated search strategies.

The slope of the AMC halo mass function (HMF) is not well constrained~\cite{Fairbairn:2017sil, Fairbairn:2017dmf,Ellis:2020gtq,Eggemeier:2019khm}. To test the dependence of our results on this slope, we re-ran the entire pipeline assuming $\mathrm{d}n/\mathrm{d}\log M \sim M^{-0.5}$, as obtained using the Press-Schechter formalism~\cite{Fairbairn:2017sil}.
Flattening the HMF increases the mean AMC mass and therefore reduces the total number of AMCs in the MW.\footnote{Note that the brightness of an individual event is only mildly dependent on the AMC mass.} Fortunately, this is partially counteracted by an increase in the encounter cross-section in Eq.~\eqref{eq:cross_section_u} which scales as $R^2 \sim M^{2/3}$. Overall, we find that the encounter rate has only a mild dependence on the slope; flattening from $-0.7$ to $-0.5$ leads to a factor of 5-10 fewer events above $1\,\mathrm{mJy}$. A recent study~\cite{Xiao:2021nkb} found an HMF slope that broadly agrees with the results of Ref.~\cite{Eggemeier:2019khm}, but with an overall shift to lower masses. This would primarily result in a decrease to the number of AMCs passing the AS cut and therefore a reduced encounter rate. 

The typical AMC density $\rho_\mathrm{AMC} \sim \delta^4$ is more strongly affected by the uncertainty in the AMC internal density profile ($\bar{\rho}_\mathrm{PL}/\bar{\rho}_\mathrm{NFW} \sim 10^5$) than by a change in $\delta$. We therefore do not expect that small variations in the distribution of $\delta$ would affect the detectability of the signal.

The production of axion miniclusters in the early Universe is a robust prediction of the post-inflationary scenario of axion cosmology~\cite{Hogan:1988mp, Kolb:1993zz, Kolb:1993hw, Kolb:1994fi, Kolb:1995bu}. The fraction of axions bound in these structures remains unclear~\cite{Eggemeier:2019khm}, but is likely to be substantial. As we show in KEVW20, if this fraction is large then direct detection efforts may be ineffective. Our results are therefore complementary to these ongoing direct searches, alongside searches for continuous radio emission from the smooth halo of axions interacting with NSs~\cite{Pshirkov:2007st, Huang:2018lxq, Hook:2018iia, Safdi:2018oeu}.

Although we have calculated the population-level distribution of signals, much work is still needed to characterize the details of each event. More concretely, the precise signal bandwidth~\cite{Battye:2019aco} and the modelling of the conversion process in realistic NS magnetospheres remain unclear. Both of these can have dramatic effects on the properties of the final signal and should be addressed in future work. We therefore emphasize that a non-detection cannot be reliably used to set upper limits on the axion parameter space. Nevertheless, this paper characterizes the unique transient nature of these interactions, and shows that current and near future radio telescopes have the sensitivity required to detect QCD axion DM.

\medskip

\begin{acknowledgments}
We thank Sebastian Baum, Gianfranco Bertone, Malte Buschmann, Matthew Lawson, David J. E. Marsh, M.C. David Marsh, Alexander Millar, Lina Necib, Ciaran O'Hare, Javier Redondo, and Ben Safdi for providing insightful comments. T.E. acknowledges support by the Vetenskapsr{\aa}det (Swedish Research Council) through contract No.  638-2013-8993 and the Oskar Klein Centre for Cosmoparticle Physics. T.E was also supported in part by the research environment grant `Detecting Axion Dark Matter In The Sky And In The Lab (AxionDM)' funded by the Swedish Research Council (VR) under Dnr 2019-02337. T.E. and C.W. are supported by the NWO through the VIDI research program ``Probing the Genesis of Dark Matter'' (680-47-5). L.V. is supported through the research program ``The Hidden Universe of Weakly Interacting Particles'' with project number 680.92.18.03 (NWO Vrije Programma), which is partly financed by the Nederlandse Organisatie voor Wetenschappelijk Onderzoek (Dutch Research Council), and acknowledges support from the European Union's Horizon 2020 research and innovation programme under the Marie Sk{\l}odowska-Curie grant agreement No.~754496 (H2020-MSCA-COFUND-2016 FELLINI). B.J.K.\ thanks the Spanish Agencia Estatal de Investigaci\'on (AEI, MICIU) for the support to the Unidad de Excelencia Mar\'ia de Maeztu Instituto de F\'isica de Cantabria, ref. MDM-2017-0765. Some of this work was carried out on the Dutch national e-infrastructure with the support of SURF Cooperative. Finally, we acknowledge the use of the Python scientific computing packages NumPy~\cite{numpy} and SciPy~\cite{scipy}, as well as the graphics environment Matplotlib~\cite{Hunter:2007}.
\end{acknowledgments}

\bibliography{main}

\begin{thebibliography}{98}%
\makeatletter
\providecommand \@ifxundefined [1]{%
 \@ifx{#1\undefined}
}%
\providecommand \@ifnum [1]{%
 \ifnum #1\expandafter \@firstoftwo
 \else \expandafter \@secondoftwo
 \fi
}%
\providecommand \@ifx [1]{%
 \ifx #1\expandafter \@firstoftwo
 \else \expandafter \@secondoftwo
 \fi
}%
\providecommand \natexlab [1]{#1}%
\providecommand \enquote  [1]{``#1''}%
\providecommand \bibnamefont  [1]{#1}%
\providecommand \bibfnamefont [1]{#1}%
\providecommand \citenamefont [1]{#1}%
\providecommand \href@noop [0]{\@secondoftwo}%
\providecommand \href [0]{\begingroup \@sanitize@url \@href}%
\providecommand \@href[1]{\@@startlink{#1}\@@href}%
\providecommand \@@href[1]{\endgroup#1\@@endlink}%
\providecommand \@sanitize@url [0]{\catcode `\\12\catcode `\$12\catcode
  `\&12\catcode `\#12\catcode `\^12\catcode `\_12\catcode `\%12\relax}%
\providecommand \@@startlink[1]{}%
\providecommand \@@endlink[0]{}%
\providecommand \url  [0]{\begingroup\@sanitize@url \@url }%
\providecommand \@url [1]{\endgroup\@href {#1}{\urlprefix }}%
\providecommand \urlprefix  [0]{URL }%
\providecommand \Eprint [0]{\href }%
\providecommand \doibase [0]{http://dx.doi.org/}%
\providecommand \selectlanguage [0]{\@gobble}%
\providecommand \bibinfo  [0]{\@secondoftwo}%
\providecommand \bibfield  [0]{\@secondoftwo}%
\providecommand \translation [1]{[#1]}%
\providecommand \BibitemOpen [0]{}%
\providecommand \bibitemStop [0]{}%
\providecommand \bibitemNoStop [0]{.\EOS\space}%
\providecommand \EOS [0]{\spacefactor3000\relax}%
\providecommand \BibitemShut  [1]{\csname bibitem#1\endcsname}%
\let\auto@bib@innerbib\@empty
\bibitem [{\citenamefont {Peccei}\ and\ \citenamefont
  {Quinn}(1977{\natexlab{a}})}]{Peccei:1977hh}%
  \BibitemOpen
  \bibfield  {author} {\bibinfo {author} {\bibfnamefont {R.~D.}\ \bibnamefont
  {Peccei}}\ and\ \bibinfo {author} {\bibfnamefont {H.~R.}\ \bibnamefont
  {Quinn}},\ }\href {\doibase 10.1103/PhysRevLett.38.1440} {\bibfield
  {journal} {\bibinfo  {journal} {Phys. Rev. Lett.}\ }\textbf {\bibinfo
  {volume} {38}},\ \bibinfo {pages} {1440} (\bibinfo {year}
  {1977}{\natexlab{a}})}\BibitemShut {NoStop}%
\bibitem [{\citenamefont {Peccei}\ and\ \citenamefont
  {Quinn}(1977{\natexlab{b}})}]{Peccei:1977ur}%
  \BibitemOpen
  \bibfield  {author} {\bibinfo {author} {\bibfnamefont {R.~D.}\ \bibnamefont
  {Peccei}}\ and\ \bibinfo {author} {\bibfnamefont {H.~R.}\ \bibnamefont
  {Quinn}},\ }\href {\doibase 10.1103/PhysRevD.16.1791} {\bibfield  {journal}
  {\bibinfo  {journal} {Phys. Rev. D}\ }\textbf {\bibinfo {volume} {16}},\
  \bibinfo {pages} {1791} (\bibinfo {year} {1977}{\natexlab{b}})}\BibitemShut
  {NoStop}%
\bibitem [{\citenamefont {Weinberg}(1978)}]{Weinberg:1977ma}%
  \BibitemOpen
  \bibfield  {author} {\bibinfo {author} {\bibfnamefont {S.}~\bibnamefont
  {Weinberg}},\ }\href {\doibase 10.1103/PhysRevLett.40.223} {\bibfield
  {journal} {\bibinfo  {journal} {Phys. Rev. Lett.}\ }\textbf {\bibinfo
  {volume} {40}},\ \bibinfo {pages} {223} (\bibinfo {year} {1978})}\BibitemShut
  {NoStop}%
\bibitem [{\citenamefont {Wilczek}(1978)}]{Wilczek:1977pj}%
  \BibitemOpen
  \bibfield  {author} {\bibinfo {author} {\bibfnamefont {F.}~\bibnamefont
  {Wilczek}},\ }\href {\doibase 10.1103/PhysRevLett.40.279} {\bibfield
  {journal} {\bibinfo  {journal} {Phys. Rev. Lett.}\ }\textbf {\bibinfo
  {volume} {40}},\ \bibinfo {pages} {279} (\bibinfo {year} {1978})}\BibitemShut
  {NoStop}%
\bibitem [{\citenamefont {Vilenkin}(1981)}]{Vilenkin:1981kz}%
  \BibitemOpen
  \bibfield  {author} {\bibinfo {author} {\bibfnamefont {A.}~\bibnamefont
  {Vilenkin}},\ }\href {\doibase 10.1103/PhysRevD.24.2082} {\bibfield
  {journal} {\bibinfo  {journal} {Phys. Rev. D}\ }\textbf {\bibinfo {volume}
  {24}},\ \bibinfo {pages} {2082} (\bibinfo {year} {1981})}\BibitemShut
  {NoStop}%
\bibitem [{\citenamefont {Abbott}\ and\ \citenamefont
  {Sikivie}(1983)}]{Abbott:1982af}%
  \BibitemOpen
  \bibfield  {author} {\bibinfo {author} {\bibfnamefont {L.~F.}\ \bibnamefont
  {Abbott}}\ and\ \bibinfo {author} {\bibfnamefont {P.}~\bibnamefont
  {Sikivie}},\ }\href {\doibase 10.1016/0370-2693(83)90638-X} {\bibfield
  {journal} {\bibinfo  {journal} {Phys. Lett. B}\ }\textbf {\bibinfo {volume}
  {120}},\ \bibinfo {pages} {133} (\bibinfo {year} {1983})}\BibitemShut
  {NoStop}%
\bibitem [{\citenamefont {Dine}\ and\ \citenamefont
  {Fischler}(1983)}]{Dine:1982ah}%
  \BibitemOpen
  \bibfield  {author} {\bibinfo {author} {\bibfnamefont {M.}~\bibnamefont
  {Dine}}\ and\ \bibinfo {author} {\bibfnamefont {W.}~\bibnamefont
  {Fischler}},\ }\href {\doibase 10.1016/0370-2693(83)90639-1} {\bibfield
  {journal} {\bibinfo  {journal} {Phys. Lett. B}\ }\textbf {\bibinfo {volume}
  {120}},\ \bibinfo {pages} {137} (\bibinfo {year} {1983})}\BibitemShut
  {NoStop}%
\bibitem [{\citenamefont {Preskill}\ \emph {et~al.}(1983)\citenamefont
  {Preskill}, \citenamefont {Wise},\ and\ \citenamefont
  {Wilczek}}]{Preskill:1982cy}%
  \BibitemOpen
  \bibfield  {author} {\bibinfo {author} {\bibfnamefont {J.}~\bibnamefont
  {Preskill}}, \bibinfo {author} {\bibfnamefont {M.~B.}\ \bibnamefont {Wise}},
  \ and\ \bibinfo {author} {\bibfnamefont {F.}~\bibnamefont {Wilczek}},\ }\href
  {\doibase 10.1016/0370-2693(83)90637-8} {\bibfield  {journal} {\bibinfo
  {journal} {Phys. Lett. B}\ }\textbf {\bibinfo {volume} {120}},\ \bibinfo
  {pages} {127} (\bibinfo {year} {1983})}\BibitemShut {NoStop}%
\bibitem [{\citenamefont {Nambu}(1960)}]{Nambu:1960tm}%
  \BibitemOpen
  \bibfield  {author} {\bibinfo {author} {\bibfnamefont {Y.}~\bibnamefont
  {Nambu}},\ }\href {\doibase 10.1103/PhysRev.117.648} {\bibfield  {journal}
  {\bibinfo  {journal} {Phys. Rev.}\ }\textbf {\bibinfo {volume} {117}},\
  \bibinfo {pages} {648} (\bibinfo {year} {1960})}\BibitemShut {NoStop}%
\bibitem [{\citenamefont {Goldstone}(1961)}]{Goldstone:1961eq}%
  \BibitemOpen
  \bibfield  {author} {\bibinfo {author} {\bibfnamefont {J.}~\bibnamefont
  {Goldstone}},\ }\href {\doibase 10.1007/BF02812722} {\bibfield  {journal}
  {\bibinfo  {journal} {Nuovo Cim.}\ }\textbf {\bibinfo {volume} {19}},\
  \bibinfo {pages} {154} (\bibinfo {year} {1961})}\BibitemShut {NoStop}%
\bibitem [{\citenamefont {Goldstone}\ \emph {et~al.}(1962)\citenamefont
  {Goldstone}, \citenamefont {Salam},\ and\ \citenamefont
  {Weinberg}}]{Goldstone:1962es}%
  \BibitemOpen
  \bibfield  {author} {\bibinfo {author} {\bibfnamefont {J.}~\bibnamefont
  {Goldstone}}, \bibinfo {author} {\bibfnamefont {A.}~\bibnamefont {Salam}}, \
  and\ \bibinfo {author} {\bibfnamefont {S.}~\bibnamefont {Weinberg}},\ }\href
  {\doibase 10.1103/PhysRev.127.965} {\bibfield  {journal} {\bibinfo  {journal}
  {Phys. Rev.}\ }\textbf {\bibinfo {volume} {127}},\ \bibinfo {pages} {965}
  (\bibinfo {year} {1962})}\BibitemShut {NoStop}%
\bibitem [{\citenamefont {Sikivie}(1983)}]{Sikivie:1983ip}%
  \BibitemOpen
  \bibfield  {author} {\bibinfo {author} {\bibfnamefont {P.}~\bibnamefont
  {Sikivie}},\ }\href {\doibase 10.1103/PhysRevLett.51.1415} {\bibfield
  {journal} {\bibinfo  {journal} {Phys. Rev. Lett.}\ }\textbf {\bibinfo
  {volume} {51}},\ \bibinfo {pages} {1415} (\bibinfo {year} {1983})},\ \bibinfo
  {note} {[Erratum: Phys.Rev.Lett. 52, 695 (1984)]}\BibitemShut {NoStop}%
\bibitem [{\citenamefont {Sikivie}(1985)}]{Sikivie:1985yu}%
  \BibitemOpen
  \bibfield  {author} {\bibinfo {author} {\bibfnamefont {P.}~\bibnamefont
  {Sikivie}},\ }\href {\doibase 10.1103/PhysRevD.36.974} {\bibfield  {journal}
  {\bibinfo  {journal} {Phys. Rev. D}\ }\textbf {\bibinfo {volume} {32}},\
  \bibinfo {pages} {2988} (\bibinfo {year} {1985})},\ \bibinfo {note}
  {[Erratum: Phys.Rev.D 36, 974 (1987)]}\BibitemShut {NoStop}%
\bibitem [{\citenamefont {Irastorza}\ and\ \citenamefont
  {Redondo}(2018)}]{Irastorza:2018dyq}%
  \BibitemOpen
  \bibfield  {author} {\bibinfo {author} {\bibfnamefont {I.~G.}\ \bibnamefont
  {Irastorza}}\ and\ \bibinfo {author} {\bibfnamefont {J.}~\bibnamefont
  {Redondo}},\ }\href {\doibase 10.1016/j.ppnp.2018.05.003} {\bibfield
  {journal} {\bibinfo  {journal} {Prog. Part. Nucl. Phys.}\ }\textbf {\bibinfo
  {volume} {102}},\ \bibinfo {pages} {89} (\bibinfo {year} {2018})},\ \Eprint
  {http://arxiv.org/abs/1801.08127} {arXiv:1801.08127 [hep-ph]} \BibitemShut
  {NoStop}%
\bibitem [{\citenamefont {Sikivie}(2021)}]{Sikivie:2020zpn}%
  \BibitemOpen
  \bibfield  {author} {\bibinfo {author} {\bibfnamefont {P.}~\bibnamefont
  {Sikivie}},\ }\href {\doibase 10.1103/RevModPhys.93.015004} {\bibfield
  {journal} {\bibinfo  {journal} {Rev. Mod. Phys.}\ }\textbf {\bibinfo {volume}
  {93}},\ \bibinfo {pages} {015004} (\bibinfo {year} {2021})},\ \Eprint
  {http://arxiv.org/abs/2003.02206} {arXiv:2003.02206 [hep-ph]} \BibitemShut
  {NoStop}%
\bibitem [{\citenamefont {Cadamuro}(2012)}]{Cadamuro:2012rm}%
  \BibitemOpen
  \bibfield  {author} {\bibinfo {author} {\bibfnamefont {D.}~\bibnamefont
  {Cadamuro}},\ }\emph {\bibinfo {title} {{Cosmological limits on axions and
  axion-like particles}}},\ \href@noop {} {Ph.D. thesis},\ \bibinfo  {school}
  {Munich U.} (\bibinfo {year} {2012}),\ \Eprint
  {http://arxiv.org/abs/1210.3196} {arXiv:1210.3196 [hep-ph]} \BibitemShut
  {NoStop}%
\bibitem [{\citenamefont {Ayala}\ \emph {et~al.}(2014)\citenamefont {Ayala},
  \citenamefont {Dom\'\i{}nguez}, \citenamefont {Giannotti}, \citenamefont
  {Mirizzi},\ and\ \citenamefont {Straniero}}]{Ayala:2014pea}%
  \BibitemOpen
  \bibfield  {author} {\bibinfo {author} {\bibfnamefont {A.}~\bibnamefont
  {Ayala}}, \bibinfo {author} {\bibfnamefont {I.}~\bibnamefont
  {Dom\'\i{}nguez}}, \bibinfo {author} {\bibfnamefont {M.}~\bibnamefont
  {Giannotti}}, \bibinfo {author} {\bibfnamefont {A.}~\bibnamefont {Mirizzi}},
  \ and\ \bibinfo {author} {\bibfnamefont {O.}~\bibnamefont {Straniero}},\
  }\href {\doibase 10.1103/PhysRevLett.113.191302} {\bibfield  {journal}
  {\bibinfo  {journal} {Phys. Rev. Lett.}\ }\textbf {\bibinfo {volume} {113}},\
  \bibinfo {pages} {191302} (\bibinfo {year} {2014})},\ \Eprint
  {http://arxiv.org/abs/1406.6053} {arXiv:1406.6053 [astro-ph.SR]} \BibitemShut
  {NoStop}%
\bibitem [{\citenamefont {Vinyoles}\ \emph {et~al.}(2015)\citenamefont
  {Vinyoles}, \citenamefont {Serenelli}, \citenamefont {Villante},
  \citenamefont {Basu}, \citenamefont {Redondo},\ and\ \citenamefont
  {Isern}}]{2015JCAP...10..015V}%
  \BibitemOpen
  \bibfield  {author} {\bibinfo {author} {\bibfnamefont {N.}~\bibnamefont
  {Vinyoles}}, \bibinfo {author} {\bibfnamefont {A.}~\bibnamefont {Serenelli}},
  \bibinfo {author} {\bibfnamefont {F.~L.}\ \bibnamefont {Villante}}, \bibinfo
  {author} {\bibfnamefont {S.}~\bibnamefont {Basu}}, \bibinfo {author}
  {\bibfnamefont {J.}~\bibnamefont {Redondo}}, \ and\ \bibinfo {author}
  {\bibfnamefont {J.}~\bibnamefont {Isern}},\ }\href {\doibase
  10.1088/1475-7516/2015/10/015} {\bibfield  {journal} {\bibinfo  {journal}
  {JCAP}\ }\textbf {\bibinfo {volume} {10}},\ \bibinfo {pages} {015} (\bibinfo
  {year} {2015})},\ \Eprint {http://arxiv.org/abs/1501.01639} {arXiv:1501.01639
  [astro-ph.SR]} \BibitemShut {NoStop}%
\bibitem [{\citenamefont {Lee}(2018)}]{Lee:2018lcj}%
  \BibitemOpen
  \bibfield  {author} {\bibinfo {author} {\bibfnamefont {J.~S.}\ \bibnamefont
  {Lee}},\ }\href@noop {} {\  (\bibinfo {year} {2018})},\ \Eprint
  {http://arxiv.org/abs/1808.10136} {arXiv:1808.10136 [hep-ph]} \BibitemShut
  {NoStop}%
\bibitem [{\citenamefont {Calore}\ \emph {et~al.}(2020)\citenamefont {Calore},
  \citenamefont {Carenza}, \citenamefont {Giannotti}, \citenamefont {Jaeckel},\
  and\ \citenamefont {Mirizzi}}]{Calore:2020tjw}%
  \BibitemOpen
  \bibfield  {author} {\bibinfo {author} {\bibfnamefont {F.}~\bibnamefont
  {Calore}}, \bibinfo {author} {\bibfnamefont {P.}~\bibnamefont {Carenza}},
  \bibinfo {author} {\bibfnamefont {M.}~\bibnamefont {Giannotti}}, \bibinfo
  {author} {\bibfnamefont {J.}~\bibnamefont {Jaeckel}}, \ and\ \bibinfo
  {author} {\bibfnamefont {A.}~\bibnamefont {Mirizzi}},\ }\href {\doibase
  10.1103/PhysRevD.102.123005} {\bibfield  {journal} {\bibinfo  {journal}
  {Phys. Rev. D}\ }\textbf {\bibinfo {volume} {102}},\ \bibinfo {pages}
  {123005} (\bibinfo {year} {2020})},\ \Eprint
  {http://arxiv.org/abs/2008.11741} {arXiv:2008.11741 [hep-ph]} \BibitemShut
  {NoStop}%
\bibitem [{\citenamefont {Pshirkov}\ and\ \citenamefont
  {Popov}(2009)}]{Pshirkov:2007st}%
  \BibitemOpen
  \bibfield  {author} {\bibinfo {author} {\bibfnamefont {M.~S.}\ \bibnamefont
  {Pshirkov}}\ and\ \bibinfo {author} {\bibfnamefont {S.~B.}\ \bibnamefont
  {Popov}},\ }\href {\doibase 10.1134/S1063776109030030} {\bibfield  {journal}
  {\bibinfo  {journal} {J. Exp. Theor. Phys.}\ }\textbf {\bibinfo {volume}
  {108}},\ \bibinfo {pages} {384} (\bibinfo {year} {2009})},\ \Eprint
  {http://arxiv.org/abs/0711.1264} {arXiv:0711.1264 [astro-ph]} \BibitemShut
  {NoStop}%
\bibitem [{\citenamefont {Huang}\ \emph {et~al.}(2018)\citenamefont {Huang},
  \citenamefont {Kadota}, \citenamefont {Sekiguchi},\ and\ \citenamefont
  {Tashiro}}]{Huang:2018lxq}%
  \BibitemOpen
  \bibfield  {author} {\bibinfo {author} {\bibfnamefont {F.~P.}\ \bibnamefont
  {Huang}}, \bibinfo {author} {\bibfnamefont {K.}~\bibnamefont {Kadota}},
  \bibinfo {author} {\bibfnamefont {T.}~\bibnamefont {Sekiguchi}}, \ and\
  \bibinfo {author} {\bibfnamefont {H.}~\bibnamefont {Tashiro}},\ }\href
  {\doibase 10.1103/PhysRevD.97.123001} {\bibfield  {journal} {\bibinfo
  {journal} {Phys. Rev. D}\ }\textbf {\bibinfo {volume} {97}},\ \bibinfo
  {pages} {123001} (\bibinfo {year} {2018})},\ \Eprint
  {http://arxiv.org/abs/1803.08230} {arXiv:1803.08230 [hep-ph]} \BibitemShut
  {NoStop}%
\bibitem [{\citenamefont {Hook}\ \emph {et~al.}(2018)\citenamefont {Hook},
  \citenamefont {Kahn}, \citenamefont {Safdi},\ and\ \citenamefont
  {Sun}}]{Hook:2018iia}%
  \BibitemOpen
  \bibfield  {author} {\bibinfo {author} {\bibfnamefont {A.}~\bibnamefont
  {Hook}}, \bibinfo {author} {\bibfnamefont {Y.}~\bibnamefont {Kahn}}, \bibinfo
  {author} {\bibfnamefont {B.~R.}\ \bibnamefont {Safdi}}, \ and\ \bibinfo
  {author} {\bibfnamefont {Z.}~\bibnamefont {Sun}},\ }\href {\doibase
  10.1103/PhysRevLett.121.241102} {\bibfield  {journal} {\bibinfo  {journal}
  {Phys. Rev. Lett.}\ }\textbf {\bibinfo {volume} {121}},\ \bibinfo {pages}
  {241102} (\bibinfo {year} {2018})},\ \Eprint
  {http://arxiv.org/abs/1804.03145} {arXiv:1804.03145 [hep-ph]} \BibitemShut
  {NoStop}%
\bibitem [{\citenamefont {Safdi}\ \emph {et~al.}(2019)\citenamefont {Safdi},
  \citenamefont {Sun},\ and\ \citenamefont {Chen}}]{Safdi:2018oeu}%
  \BibitemOpen
  \bibfield  {author} {\bibinfo {author} {\bibfnamefont {B.~R.}\ \bibnamefont
  {Safdi}}, \bibinfo {author} {\bibfnamefont {Z.}~\bibnamefont {Sun}}, \ and\
  \bibinfo {author} {\bibfnamefont {A.~Y.}\ \bibnamefont {Chen}},\ }\href
  {\doibase 10.1103/PhysRevD.99.123021} {\bibfield  {journal} {\bibinfo
  {journal} {Phys. Rev. D}\ }\textbf {\bibinfo {volume} {99}},\ \bibinfo
  {pages} {123021} (\bibinfo {year} {2019})},\ \Eprint
  {http://arxiv.org/abs/1811.01020} {arXiv:1811.01020 [astro-ph.CO]}
  \BibitemShut {NoStop}%
\bibitem [{\citenamefont {Edwards}\ \emph {et~al.}(2020)\citenamefont
  {Edwards}, \citenamefont {Chianese}, \citenamefont {Kavanagh}, \citenamefont
  {Nissanke},\ and\ \citenamefont {Weniger}}]{Edwards:2019tzf}%
  \BibitemOpen
  \bibfield  {author} {\bibinfo {author} {\bibfnamefont {T.~D.~P.}\
  \bibnamefont {Edwards}}, \bibinfo {author} {\bibfnamefont {M.}~\bibnamefont
  {Chianese}}, \bibinfo {author} {\bibfnamefont {B.~J.}\ \bibnamefont
  {Kavanagh}}, \bibinfo {author} {\bibfnamefont {S.~M.}\ \bibnamefont
  {Nissanke}}, \ and\ \bibinfo {author} {\bibfnamefont {C.}~\bibnamefont
  {Weniger}},\ }\href {\doibase 10.1103/PhysRevLett.124.161101} {\bibfield
  {journal} {\bibinfo  {journal} {Phys. Rev. Lett.}\ }\textbf {\bibinfo
  {volume} {124}},\ \bibinfo {pages} {161101} (\bibinfo {year} {2020})},\
  \Eprint {http://arxiv.org/abs/1905.04686} {arXiv:1905.04686 [hep-ph]}
  \BibitemShut {NoStop}%
\bibitem [{\citenamefont {Foster}\ \emph {et~al.}(2020)\citenamefont {Foster},
  \citenamefont {Kahn}, \citenamefont {Macias}, \citenamefont {Sun},
  \citenamefont {Eatough}, \citenamefont {Kondratiev}, \citenamefont {Peters},
  \citenamefont {Weniger},\ and\ \citenamefont {Safdi}}]{Foster:2020pgt}%
  \BibitemOpen
  \bibfield  {author} {\bibinfo {author} {\bibfnamefont {J.~W.}\ \bibnamefont
  {Foster}}, \bibinfo {author} {\bibfnamefont {Y.}~\bibnamefont {Kahn}},
  \bibinfo {author} {\bibfnamefont {O.}~\bibnamefont {Macias}}, \bibinfo
  {author} {\bibfnamefont {Z.}~\bibnamefont {Sun}}, \bibinfo {author}
  {\bibfnamefont {R.~P.}\ \bibnamefont {Eatough}}, \bibinfo {author}
  {\bibfnamefont {V.~I.}\ \bibnamefont {Kondratiev}}, \bibinfo {author}
  {\bibfnamefont {W.~M.}\ \bibnamefont {Peters}}, \bibinfo {author}
  {\bibfnamefont {C.}~\bibnamefont {Weniger}}, \ and\ \bibinfo {author}
  {\bibfnamefont {B.~R.}\ \bibnamefont {Safdi}},\ }\href {\doibase
  10.1103/PhysRevLett.125.171301} {\bibfield  {journal} {\bibinfo  {journal}
  {Phys. Rev. Lett.}\ }\textbf {\bibinfo {volume} {125}},\ \bibinfo {pages}
  {171301} (\bibinfo {year} {2020})},\ \Eprint
  {http://arxiv.org/abs/2004.00011} {arXiv:2004.00011 [astro-ph.CO]}
  \BibitemShut {NoStop}%
\bibitem [{\citenamefont {Darling}(2020{\natexlab{a}})}]{Darling:2020uyo}%
  \BibitemOpen
  \bibfield  {author} {\bibinfo {author} {\bibfnamefont {J.}~\bibnamefont
  {Darling}},\ }\href {\doibase 10.3847/2041-8213/abb23f} {\bibfield  {journal}
  {\bibinfo  {journal} {Astrophys. J. Lett.}\ }\textbf {\bibinfo {volume}
  {900}},\ \bibinfo {pages} {L28} (\bibinfo {year} {2020}{\natexlab{a}})},\
  \Eprint {http://arxiv.org/abs/2008.11188} {arXiv:2008.11188 [astro-ph.CO]}
  \BibitemShut {NoStop}%
\bibitem [{\citenamefont {Hogan}\ and\ \citenamefont
  {Rees}(1988)}]{Hogan:1988mp}%
  \BibitemOpen
  \bibfield  {author} {\bibinfo {author} {\bibfnamefont {C.~J.}\ \bibnamefont
  {Hogan}}\ and\ \bibinfo {author} {\bibfnamefont {M.~J.}\ \bibnamefont
  {Rees}},\ }\href {\doibase 10.1016/0370-2693(88)91655-3} {\bibfield
  {journal} {\bibinfo  {journal} {Phys. Lett. B}\ }\textbf {\bibinfo {volume}
  {205}},\ \bibinfo {pages} {228} (\bibinfo {year} {1988})}\BibitemShut
  {NoStop}%
\bibitem [{\citenamefont {Kolb}\ and\ \citenamefont
  {Tkachev}(1993)}]{Kolb:1993zz}%
  \BibitemOpen
  \bibfield  {author} {\bibinfo {author} {\bibfnamefont {E.~W.}\ \bibnamefont
  {Kolb}}\ and\ \bibinfo {author} {\bibfnamefont {I.~I.}\ \bibnamefont
  {Tkachev}},\ }\href {\doibase 10.1103/PhysRevLett.71.3051} {\bibfield
  {journal} {\bibinfo  {journal} {Phys. Rev. Lett.}\ }\textbf {\bibinfo
  {volume} {71}},\ \bibinfo {pages} {3051} (\bibinfo {year} {1993})},\ \Eprint
  {http://arxiv.org/abs/hep-ph/9303313} {arXiv:hep-ph/9303313} \BibitemShut
  {NoStop}%
\bibitem [{\citenamefont {Kolb}\ and\ \citenamefont
  {Tkachev}(1994{\natexlab{a}})}]{Kolb:1993hw}%
  \BibitemOpen
  \bibfield  {author} {\bibinfo {author} {\bibfnamefont {E.~W.}\ \bibnamefont
  {Kolb}}\ and\ \bibinfo {author} {\bibfnamefont {I.~I.}\ \bibnamefont
  {Tkachev}},\ }\href {\doibase 10.1103/PhysRevD.49.5040} {\bibfield  {journal}
  {\bibinfo  {journal} {Phys. Rev. D}\ }\textbf {\bibinfo {volume} {49}},\
  \bibinfo {pages} {5040} (\bibinfo {year} {1994}{\natexlab{a}})},\ \Eprint
  {http://arxiv.org/abs/astro-ph/9311037} {arXiv:astro-ph/9311037} \BibitemShut
  {NoStop}%
\bibitem [{\citenamefont {Kolb}\ and\ \citenamefont
  {Tkachev}(1994{\natexlab{b}})}]{Kolb:1994fi}%
  \BibitemOpen
  \bibfield  {author} {\bibinfo {author} {\bibfnamefont {E.~W.}\ \bibnamefont
  {Kolb}}\ and\ \bibinfo {author} {\bibfnamefont {I.~I.}\ \bibnamefont
  {Tkachev}},\ }\href {\doibase 10.1103/PhysRevD.50.769} {\bibfield  {journal}
  {\bibinfo  {journal} {Phys. Rev. D}\ }\textbf {\bibinfo {volume} {50}},\
  \bibinfo {pages} {769} (\bibinfo {year} {1994}{\natexlab{b}})},\ \Eprint
  {http://arxiv.org/abs/astro-ph/9403011} {arXiv:astro-ph/9403011} \BibitemShut
  {NoStop}%
\bibitem [{\citenamefont {Kolb}\ and\ \citenamefont
  {Tkachev}(1996)}]{Kolb:1995bu}%
  \BibitemOpen
  \bibfield  {author} {\bibinfo {author} {\bibfnamefont {E.~W.}\ \bibnamefont
  {Kolb}}\ and\ \bibinfo {author} {\bibfnamefont {I.~I.}\ \bibnamefont
  {Tkachev}},\ }\href {\doibase 10.1086/309962} {\bibfield  {journal} {\bibinfo
   {journal} {Astrophys. J. Lett.}\ }\textbf {\bibinfo {volume} {460}},\
  \bibinfo {pages} {L25} (\bibinfo {year} {1996})},\ \Eprint
  {http://arxiv.org/abs/astro-ph/9510043} {arXiv:astro-ph/9510043} \BibitemShut
  {NoStop}%
\bibitem [{\citenamefont {Eggemeier}\ \emph {et~al.}(2020)\citenamefont
  {Eggemeier}, \citenamefont {Redondo}, \citenamefont {Dolag}, \citenamefont
  {Niemeyer},\ and\ \citenamefont {Vaquero}}]{Eggemeier:2019khm}%
  \BibitemOpen
  \bibfield  {author} {\bibinfo {author} {\bibfnamefont {B.}~\bibnamefont
  {Eggemeier}}, \bibinfo {author} {\bibfnamefont {J.}~\bibnamefont {Redondo}},
  \bibinfo {author} {\bibfnamefont {K.}~\bibnamefont {Dolag}}, \bibinfo
  {author} {\bibfnamefont {J.~C.}\ \bibnamefont {Niemeyer}}, \ and\ \bibinfo
  {author} {\bibfnamefont {A.}~\bibnamefont {Vaquero}},\ }\href {\doibase
  10.1103/PhysRevLett.125.041301} {\bibfield  {journal} {\bibinfo  {journal}
  {Phys. Rev. Lett.}\ }\textbf {\bibinfo {volume} {125}},\ \bibinfo {pages}
  {041301} (\bibinfo {year} {2020})},\ \Eprint
  {http://arxiv.org/abs/1911.09417} {arXiv:1911.09417 [astro-ph.CO]}
  \BibitemShut {NoStop}%
\bibitem [{\citenamefont {Goldreich}\ and\ \citenamefont
  {Julian}(1969)}]{Goldreich:1969sb}%
  \BibitemOpen
  \bibfield  {author} {\bibinfo {author} {\bibfnamefont {P.}~\bibnamefont
  {Goldreich}}\ and\ \bibinfo {author} {\bibfnamefont {W.~H.}\ \bibnamefont
  {Julian}},\ }\href {\doibase 10.1086/150119} {\bibfield  {journal} {\bibinfo
  {journal} {Astrophys. J.}\ }\textbf {\bibinfo {volume} {157}},\ \bibinfo
  {pages} {869} (\bibinfo {year} {1969})}\BibitemShut {NoStop}%
\bibitem [{\citenamefont {Phinney}\ and\ \citenamefont
  {Kulkarni}(1994)}]{Phinney:1994gf}%
  \BibitemOpen
  \bibfield  {author} {\bibinfo {author} {\bibfnamefont {E.~S.}\ \bibnamefont
  {Phinney}}\ and\ \bibinfo {author} {\bibfnamefont {S.~R.}\ \bibnamefont
  {Kulkarni}},\ }\href {\doibase 10.1146/annurev.aa.32.090194.003111}
  {\bibfield  {journal} {\bibinfo  {journal} {Ann. Rev. Astron. Astrophys.}\
  }\textbf {\bibinfo {volume} {32}},\ \bibinfo {pages} {591} (\bibinfo {year}
  {1994})}\BibitemShut {NoStop}%
\bibitem [{\citenamefont {Vaquero}\ \emph {et~al.}(2019)\citenamefont
  {Vaquero}, \citenamefont {Redondo},\ and\ \citenamefont
  {Stadler}}]{Vaquero:2018tib}%
  \BibitemOpen
  \bibfield  {author} {\bibinfo {author} {\bibfnamefont {A.}~\bibnamefont
  {Vaquero}}, \bibinfo {author} {\bibfnamefont {J.}~\bibnamefont {Redondo}}, \
  and\ \bibinfo {author} {\bibfnamefont {J.}~\bibnamefont {Stadler}},\ }\href
  {\doibase 10.1088/1475-7516/2019/04/012} {\bibfield  {journal} {\bibinfo
  {journal} {JCAP}\ }\textbf {\bibinfo {volume} {04}},\ \bibinfo {pages} {012}
  (\bibinfo {year} {2019})},\ \Eprint {http://arxiv.org/abs/1809.09241}
  {arXiv:1809.09241 [astro-ph.CO]} \BibitemShut {NoStop}%
\bibitem [{\citenamefont {Zurek}\ \emph {et~al.}(2007)\citenamefont {Zurek},
  \citenamefont {Hogan},\ and\ \citenamefont {Quinn}}]{Zurek:2006sy}%
  \BibitemOpen
  \bibfield  {author} {\bibinfo {author} {\bibfnamefont {K.~M.}\ \bibnamefont
  {Zurek}}, \bibinfo {author} {\bibfnamefont {C.~J.}\ \bibnamefont {Hogan}}, \
  and\ \bibinfo {author} {\bibfnamefont {T.~R.}\ \bibnamefont {Quinn}},\ }\href
  {\doibase 10.1103/PhysRevD.75.043511} {\bibfield  {journal} {\bibinfo
  {journal} {Phys. Rev. D}\ }\textbf {\bibinfo {volume} {75}},\ \bibinfo
  {pages} {043511} (\bibinfo {year} {2007})},\ \Eprint
  {http://arxiv.org/abs/astro-ph/0607341} {arXiv:astro-ph/0607341} \BibitemShut
  {NoStop}%
\bibitem [{\citenamefont {Buschmann}\ \emph {et~al.}(2020)\citenamefont
  {Buschmann}, \citenamefont {Foster},\ and\ \citenamefont
  {Safdi}}]{Buschmann:2019icd}%
  \BibitemOpen
  \bibfield  {author} {\bibinfo {author} {\bibfnamefont {M.}~\bibnamefont
  {Buschmann}}, \bibinfo {author} {\bibfnamefont {J.~W.}\ \bibnamefont
  {Foster}}, \ and\ \bibinfo {author} {\bibfnamefont {B.~R.}\ \bibnamefont
  {Safdi}},\ }\href {\doibase 10.1103/PhysRevLett.124.161103} {\bibfield
  {journal} {\bibinfo  {journal} {Phys. Rev. Lett.}\ }\textbf {\bibinfo
  {volume} {124}},\ \bibinfo {pages} {161103} (\bibinfo {year} {2020})},\
  \Eprint {http://arxiv.org/abs/1906.00967} {arXiv:1906.00967 [astro-ph.CO]}
  \BibitemShut {NoStop}%
\bibitem [{\citenamefont {Visinelli}\ and\ \citenamefont
  {Redondo}(2020)}]{Visinelli:2018wza}%
  \BibitemOpen
  \bibfield  {author} {\bibinfo {author} {\bibfnamefont {L.}~\bibnamefont
  {Visinelli}}\ and\ \bibinfo {author} {\bibfnamefont {J.}~\bibnamefont
  {Redondo}},\ }\href {\doibase 10.1103/PhysRevD.101.023008} {\bibfield
  {journal} {\bibinfo  {journal} {Phys. Rev. D}\ }\textbf {\bibinfo {volume}
  {101}},\ \bibinfo {pages} {023008} (\bibinfo {year} {2020})},\ \Eprint
  {http://arxiv.org/abs/1808.01879} {arXiv:1808.01879 [astro-ph.CO]}
  \BibitemShut {NoStop}%
\bibitem [{\citenamefont {Kavanagh}\ \emph
  {et~al.}(2020{\natexlab{a}})\citenamefont {Kavanagh}, \citenamefont
  {Edwards}, \citenamefont {Visinelli},\ and\ \citenamefont
  {Weniger}}]{Kavanagh:2020gcy}%
  \BibitemOpen
  \bibfield  {author} {\bibinfo {author} {\bibfnamefont {B.~J.}\ \bibnamefont
  {Kavanagh}}, \bibinfo {author} {\bibfnamefont {T.~D.~P.}\ \bibnamefont
  {Edwards}}, \bibinfo {author} {\bibfnamefont {L.}~\bibnamefont {Visinelli}},
  \ and\ \bibinfo {author} {\bibfnamefont {C.}~\bibnamefont {Weniger}},\
  }\href@noop {} {\  (\bibinfo {year} {2020}{\natexlab{a}})},\ \Eprint
  {http://arxiv.org/abs/2011.05377} {arXiv:2011.05377 [astro-ph.GA]}
  \BibitemShut {NoStop}%
\bibitem [{\citenamefont {Kim}(1979)}]{Kim:1979if}%
  \BibitemOpen
  \bibfield  {author} {\bibinfo {author} {\bibfnamefont {J.~E.}\ \bibnamefont
  {Kim}},\ }\href {\doibase 10.1103/PhysRevLett.43.103} {\bibfield  {journal}
  {\bibinfo  {journal} {Phys. Rev. Lett.}\ }\textbf {\bibinfo {volume} {43}},\
  \bibinfo {pages} {103} (\bibinfo {year} {1979})}\BibitemShut {NoStop}%
\bibitem [{\citenamefont {Shifman}\ \emph {et~al.}(1980)\citenamefont
  {Shifman}, \citenamefont {Vainshtein},\ and\ \citenamefont
  {Zakharov}}]{Shifman:1979if}%
  \BibitemOpen
  \bibfield  {author} {\bibinfo {author} {\bibfnamefont {M.~A.}\ \bibnamefont
  {Shifman}}, \bibinfo {author} {\bibfnamefont {A.~I.}\ \bibnamefont
  {Vainshtein}}, \ and\ \bibinfo {author} {\bibfnamefont {V.~I.}\ \bibnamefont
  {Zakharov}},\ }\href {\doibase 10.1016/0550-3213(80)90209-6} {\bibfield
  {journal} {\bibinfo  {journal} {Nucl. Phys. B}\ }\textbf {\bibinfo {volume}
  {166}},\ \bibinfo {pages} {493} (\bibinfo {year} {1980})}\BibitemShut
  {NoStop}%
\bibitem [{\citenamefont {Klaer}\ and\ \citenamefont
  {Moore}(2017)}]{Klaer:2017ond}%
  \BibitemOpen
  \bibfield  {author} {\bibinfo {author} {\bibfnamefont {V.~B.~.}\ \bibnamefont
  {Klaer}}\ and\ \bibinfo {author} {\bibfnamefont {G.~D.}\ \bibnamefont
  {Moore}},\ }\href {\doibase 10.1088/1475-7516/2017/11/049} {\bibfield
  {journal} {\bibinfo  {journal} {JCAP}\ }\textbf {\bibinfo {volume} {11}},\
  \bibinfo {pages} {049} (\bibinfo {year} {2017})},\ \Eprint
  {http://arxiv.org/abs/1708.07521} {arXiv:1708.07521 [hep-ph]} \BibitemShut
  {NoStop}%
\bibitem [{\citenamefont {Kavanagh}\ \emph
  {et~al.}(2020{\natexlab{b}})\citenamefont {Kavanagh}, \citenamefont
  {Edwards},\ and\ \citenamefont {Visinelli}}]{AMC_code}%
  \BibitemOpen
  \bibfield  {author} {\bibinfo {author} {\bibfnamefont {B.~J.}\ \bibnamefont
  {Kavanagh}}, \bibinfo {author} {\bibfnamefont {T.~D.~P.}\ \bibnamefont
  {Edwards}}, \ and\ \bibinfo {author} {\bibfnamefont {L.}~\bibnamefont
  {Visinelli}},\ }\href {\doibase 10.5281/zenodo.4006128} {\enquote {\bibinfo
  {title} {axion-miniclusters [computer software]},}\ }\bibinfo {howpublished}
  {\href{https://github.com/bradkav/axion-miniclusters/}{github.com/bradkav/axion-miniclusters},
  archived at
  \href{https://doi.org/10.5281/zenodo.4006128}{DOI:10.5281/zenodo.4006128}}
  (\bibinfo {year} {2020}{\natexlab{b}})\BibitemShut {NoStop}%
\bibitem [{\citenamefont {Fairbairn}\ \emph {et~al.}(2018)\citenamefont
  {Fairbairn}, \citenamefont {Marsh}, \citenamefont {Quevillon},\ and\
  \citenamefont {Rozier}}]{Fairbairn:2017sil}%
  \BibitemOpen
  \bibfield  {author} {\bibinfo {author} {\bibfnamefont {M.}~\bibnamefont
  {Fairbairn}}, \bibinfo {author} {\bibfnamefont {D.~J.~E.}\ \bibnamefont
  {Marsh}}, \bibinfo {author} {\bibfnamefont {J.}~\bibnamefont {Quevillon}}, \
  and\ \bibinfo {author} {\bibfnamefont {S.}~\bibnamefont {Rozier}},\ }\href
  {\doibase 10.1103/PhysRevD.97.083502} {\bibfield  {journal} {\bibinfo
  {journal} {Phys. Rev. D}\ }\textbf {\bibinfo {volume} {97}},\ \bibinfo
  {pages} {083502} (\bibinfo {year} {2018})},\ \Eprint
  {http://arxiv.org/abs/1707.03310} {arXiv:1707.03310 [astro-ph.CO]}
  \BibitemShut {NoStop}%
\bibitem [{\citenamefont {Fairbairn}\ \emph {et~al.}(2017)\citenamefont
  {Fairbairn}, \citenamefont {Marsh},\ and\ \citenamefont
  {Quevillon}}]{Fairbairn:2017dmf}%
  \BibitemOpen
  \bibfield  {author} {\bibinfo {author} {\bibfnamefont {M.}~\bibnamefont
  {Fairbairn}}, \bibinfo {author} {\bibfnamefont {D.~J.~E.}\ \bibnamefont
  {Marsh}}, \ and\ \bibinfo {author} {\bibfnamefont {J.}~\bibnamefont
  {Quevillon}},\ }\href {\doibase 10.1103/PhysRevLett.119.021101} {\bibfield
  {journal} {\bibinfo  {journal} {Phys. Rev. Lett.}\ }\textbf {\bibinfo
  {volume} {119}},\ \bibinfo {pages} {021101} (\bibinfo {year} {2017})},\
  \Eprint {http://arxiv.org/abs/1701.04787} {arXiv:1701.04787 [astro-ph.CO]}
  \BibitemShut {NoStop}%
\bibitem [{\citenamefont {Ellis}\ \emph {et~al.}(2021)\citenamefont {Ellis},
  \citenamefont {Marsh},\ and\ \citenamefont {Behrens}}]{Ellis:2020gtq}%
  \BibitemOpen
  \bibfield  {author} {\bibinfo {author} {\bibfnamefont {D.}~\bibnamefont
  {Ellis}}, \bibinfo {author} {\bibfnamefont {D.~J.~E.}\ \bibnamefont {Marsh}},
  \ and\ \bibinfo {author} {\bibfnamefont {C.}~\bibnamefont {Behrens}},\ }\href
  {\doibase 10.1103/PhysRevD.103.083525} {\bibfield  {journal} {\bibinfo
  {journal} {Phys. Rev. D}\ }\textbf {\bibinfo {volume} {103}},\ \bibinfo
  {pages} {083525} (\bibinfo {year} {2021})},\ \Eprint
  {http://arxiv.org/abs/2006.08637} {arXiv:2006.08637 [astro-ph.CO]}
  \BibitemShut {NoStop}%
\bibitem [{\citenamefont {Zhao}\ \emph {et~al.}(2005)\citenamefont {Zhao},
  \citenamefont {Taylor}, \citenamefont {Silk},\ and\ \citenamefont
  {Hooper}}]{Zhao:2005py}%
  \BibitemOpen
  \bibfield  {author} {\bibinfo {author} {\bibfnamefont {H.-S.}\ \bibnamefont
  {Zhao}}, \bibinfo {author} {\bibfnamefont {J.}~\bibnamefont {Taylor}},
  \bibinfo {author} {\bibfnamefont {J.}~\bibnamefont {Silk}}, \ and\ \bibinfo
  {author} {\bibfnamefont {D.}~\bibnamefont {Hooper}},\ }\href@noop {} {\
  (\bibinfo {year} {2005})},\ \Eprint {http://arxiv.org/abs/astro-ph/0502049}
  {arXiv:astro-ph/0502049} \BibitemShut {NoStop}%
\bibitem [{\citenamefont {Tinyakov}\ \emph {et~al.}(2016)\citenamefont
  {Tinyakov}, \citenamefont {Tkachev},\ and\ \citenamefont
  {Zioutas}}]{Tinyakov:2015cgg}%
  \BibitemOpen
  \bibfield  {author} {\bibinfo {author} {\bibfnamefont {P.}~\bibnamefont
  {Tinyakov}}, \bibinfo {author} {\bibfnamefont {I.}~\bibnamefont {Tkachev}}, \
  and\ \bibinfo {author} {\bibfnamefont {K.}~\bibnamefont {Zioutas}},\ }\href
  {\doibase 10.1088/1475-7516/2016/01/035} {\bibfield  {journal} {\bibinfo
  {journal} {JCAP}\ }\textbf {\bibinfo {volume} {01}},\ \bibinfo {pages} {035}
  (\bibinfo {year} {2016})},\ \Eprint {http://arxiv.org/abs/1512.02884}
  {arXiv:1512.02884 [astro-ph.CO]} \BibitemShut {NoStop}%
\bibitem [{\citenamefont {Berezinsky}\ \emph {et~al.}(2013)\citenamefont
  {Berezinsky}, \citenamefont {Dokuchaev},\ and\ \citenamefont
  {Eroshenko}}]{Berezinsky:2013fxa}%
  \BibitemOpen
  \bibfield  {author} {\bibinfo {author} {\bibfnamefont {V.~S.}\ \bibnamefont
  {Berezinsky}}, \bibinfo {author} {\bibfnamefont {V.~I.}\ \bibnamefont
  {Dokuchaev}}, \ and\ \bibinfo {author} {\bibfnamefont {Y.~N.}\ \bibnamefont
  {Eroshenko}},\ }\href {\doibase 10.1088/1475-7516/2013/11/059} {\bibfield
  {journal} {\bibinfo  {journal} {JCAP}\ }\textbf {\bibinfo {volume} {11}},\
  \bibinfo {pages} {059} (\bibinfo {year} {2013})},\ \Eprint
  {http://arxiv.org/abs/1308.6742} {arXiv:1308.6742 [astro-ph.CO]} \BibitemShut
  {NoStop}%
\bibitem [{\citenamefont {Dokuchaev}\ \emph {et~al.}(2017)\citenamefont
  {Dokuchaev}, \citenamefont {Eroshenko},\ and\ \citenamefont
  {Tkachev}}]{2017JETP..125..434D}%
  \BibitemOpen
  \bibfield  {author} {\bibinfo {author} {\bibfnamefont {V.~I.}\ \bibnamefont
  {Dokuchaev}}, \bibinfo {author} {\bibfnamefont {Y.~N.}\ \bibnamefont
  {Eroshenko}}, \ and\ \bibinfo {author} {\bibfnamefont {I.~I.}\ \bibnamefont
  {Tkachev}},\ }\href {\doibase 10.1134/S1063776117080039} {\bibfield
  {journal} {\bibinfo  {journal} {J. Exp. Theor. Phys.}\ }\textbf {\bibinfo
  {volume} {125}},\ \bibinfo {pages} {434} (\bibinfo {year} {2017})},\ \Eprint
  {http://arxiv.org/abs/1710.09586} {arXiv:1710.09586 [astro-ph.GA]}
  \BibitemShut {NoStop}%
\bibitem [{\citenamefont {Navarro}\ \emph {et~al.}(1996)\citenamefont
  {Navarro}, \citenamefont {Frenk},\ and\ \citenamefont
  {White}}]{Navarro:1995iw}%
  \BibitemOpen
  \bibfield  {author} {\bibinfo {author} {\bibfnamefont {J.~F.}\ \bibnamefont
  {Navarro}}, \bibinfo {author} {\bibfnamefont {C.~S.}\ \bibnamefont {Frenk}},
  \ and\ \bibinfo {author} {\bibfnamefont {S.~D.~M.}\ \bibnamefont {White}},\
  }\href {\doibase 10.1086/177173} {\bibfield  {journal} {\bibinfo  {journal}
  {Astrophys. J.}\ }\textbf {\bibinfo {volume} {462}},\ \bibinfo {pages} {563}
  (\bibinfo {year} {1996})},\ \Eprint {http://arxiv.org/abs/astro-ph/9508025}
  {arXiv:astro-ph/9508025} \BibitemShut {NoStop}%
\bibitem [{\citenamefont {Seidel}\ and\ \citenamefont
  {Suen}(1994)}]{Seidel:1993zk}%
  \BibitemOpen
  \bibfield  {author} {\bibinfo {author} {\bibfnamefont {E.}~\bibnamefont
  {Seidel}}\ and\ \bibinfo {author} {\bibfnamefont {W.-M.}\ \bibnamefont
  {Suen}},\ }\href {\doibase 10.1103/PhysRevLett.72.2516} {\bibfield  {journal}
  {\bibinfo  {journal} {Phys. Rev. Lett.}\ }\textbf {\bibinfo {volume} {72}},\
  \bibinfo {pages} {2516} (\bibinfo {year} {1994})},\ \Eprint
  {http://arxiv.org/abs/gr-qc/9309015} {arXiv:gr-qc/9309015} \BibitemShut
  {NoStop}%
\bibitem [{\citenamefont {Levkov}\ \emph {et~al.}(2018)\citenamefont {Levkov},
  \citenamefont {Panin},\ and\ \citenamefont {Tkachev}}]{Levkov:2018kau}%
  \BibitemOpen
  \bibfield  {author} {\bibinfo {author} {\bibfnamefont {D.~G.}\ \bibnamefont
  {Levkov}}, \bibinfo {author} {\bibfnamefont {A.~G.}\ \bibnamefont {Panin}}, \
  and\ \bibinfo {author} {\bibfnamefont {I.~I.}\ \bibnamefont {Tkachev}},\
  }\href {\doibase 10.1103/PhysRevLett.121.151301} {\bibfield  {journal}
  {\bibinfo  {journal} {Phys. Rev. Lett.}\ }\textbf {\bibinfo {volume} {121}},\
  \bibinfo {pages} {151301} (\bibinfo {year} {2018})},\ \Eprint
  {http://arxiv.org/abs/1804.05857} {arXiv:1804.05857 [astro-ph.CO]}
  \BibitemShut {NoStop}%
\bibitem [{\citenamefont {Eggemeier}\ and\ \citenamefont
  {Niemeyer}(2019)}]{Eggemeier:2019jsu}%
  \BibitemOpen
  \bibfield  {author} {\bibinfo {author} {\bibfnamefont {B.}~\bibnamefont
  {Eggemeier}}\ and\ \bibinfo {author} {\bibfnamefont {J.~C.}\ \bibnamefont
  {Niemeyer}},\ }\href {\doibase 10.1103/PhysRevD.100.063528} {\bibfield
  {journal} {\bibinfo  {journal} {Phys. Rev. D}\ }\textbf {\bibinfo {volume}
  {100}},\ \bibinfo {pages} {063528} (\bibinfo {year} {2019})},\ \Eprint
  {http://arxiv.org/abs/1906.01348} {arXiv:1906.01348 [astro-ph.CO]}
  \BibitemShut {NoStop}%
\bibitem [{\citenamefont {Chen}\ \emph {et~al.}(2020)\citenamefont {Chen},
  \citenamefont {Du}, \citenamefont {Lentz}, \citenamefont {Marsh},\ and\
  \citenamefont {Niemeyer}}]{Chen:2020cef}%
  \BibitemOpen
  \bibfield  {author} {\bibinfo {author} {\bibfnamefont {J.}~\bibnamefont
  {Chen}}, \bibinfo {author} {\bibfnamefont {X.}~\bibnamefont {Du}}, \bibinfo
  {author} {\bibfnamefont {E.~W.}\ \bibnamefont {Lentz}}, \bibinfo {author}
  {\bibfnamefont {D.~J.~E.}\ \bibnamefont {Marsh}}, \ and\ \bibinfo {author}
  {\bibfnamefont {J.~C.}\ \bibnamefont {Niemeyer}},\ }\href@noop {} {\
  (\bibinfo {year} {2020})},\ \Eprint {http://arxiv.org/abs/2011.01333}
  {arXiv:2011.01333 [astro-ph.CO]} \BibitemShut {NoStop}%
\bibitem [{\citenamefont {{Binney}}\ and\ \citenamefont
  {{Tremaine}}(2008)}]{BinneyTremaine:2008}%
  \BibitemOpen
  \bibfield  {author} {\bibinfo {author} {\bibfnamefont {J.}~\bibnamefont
  {{Binney}}}\ and\ \bibinfo {author} {\bibfnamefont {S.}~\bibnamefont
  {{Tremaine}}},\ }\href@noop {} {\emph {\bibinfo {title} {Galactic Dynamics:
  Second Edition, by James Binney and Scott Tremaine.~ISBN 978-0-691-13026-2
  (HB).~Published by Princeton University Press, Princeton, NJ USA, 2008.}}}\
  (\bibinfo  {publisher} {Princeton University Press},\ \bibinfo {year}
  {2008})\BibitemShut {NoStop}%
\bibitem [{\citenamefont {Buckley}\ \emph {et~al.}(2021)\citenamefont
  {Buckley}, \citenamefont {Dev}, \citenamefont {Ferrer},\ and\ \citenamefont
  {Huang}}]{Buckley:2020fmh}%
  \BibitemOpen
  \bibfield  {author} {\bibinfo {author} {\bibfnamefont {J.~H.}\ \bibnamefont
  {Buckley}}, \bibinfo {author} {\bibfnamefont {P.~S.~B.}\ \bibnamefont {Dev}},
  \bibinfo {author} {\bibfnamefont {F.}~\bibnamefont {Ferrer}}, \ and\ \bibinfo
  {author} {\bibfnamefont {F.~P.}\ \bibnamefont {Huang}},\ }\href {\doibase
  10.1103/PhysRevD.103.043015} {\bibfield  {journal} {\bibinfo  {journal}
  {Phys. Rev. D}\ }\textbf {\bibinfo {volume} {103}},\ \bibinfo {pages}
  {043015} (\bibinfo {year} {2021})},\ \Eprint
  {http://arxiv.org/abs/2004.06486} {arXiv:2004.06486 [astro-ph.HE]}
  \BibitemShut {NoStop}%
\bibitem [{\citenamefont {Prabhu}\ and\ \citenamefont
  {Rapidis}(2020)}]{Prabhu:2020yif}%
  \BibitemOpen
  \bibfield  {author} {\bibinfo {author} {\bibfnamefont {A.}~\bibnamefont
  {Prabhu}}\ and\ \bibinfo {author} {\bibfnamefont {N.~M.}\ \bibnamefont
  {Rapidis}},\ }\href {\doibase 10.1088/1475-7516/2020/10/054} {\bibfield
  {journal} {\bibinfo  {journal} {JCAP}\ }\textbf {\bibinfo {volume} {10}},\
  \bibinfo {pages} {054} (\bibinfo {year} {2020})},\ \Eprint
  {http://arxiv.org/abs/2005.03700} {arXiv:2005.03700 [astro-ph.CO]}
  \BibitemShut {NoStop}%
\bibitem [{\citenamefont {Ofek}(2009)}]{Ofek:2009wt}%
  \BibitemOpen
  \bibfield  {author} {\bibinfo {author} {\bibfnamefont {E.~O.}\ \bibnamefont
  {Ofek}},\ }\href {\doibase 10.1086/605389} {\bibfield  {journal} {\bibinfo
  {journal} {Publ. Astron. Soc. Pac.}\ }\textbf {\bibinfo {volume} {121}},\
  \bibinfo {pages} {814} (\bibinfo {year} {2009})},\ \Eprint
  {http://arxiv.org/abs/0910.3684} {arXiv:0910.3684 [astro-ph.GA]} \BibitemShut
  {NoStop}%
\bibitem [{\citenamefont {Sartore}\ \emph {et~al.}(2010)\citenamefont
  {Sartore}, \citenamefont {Ripamonti}, \citenamefont {Treves},\ and\
  \citenamefont {Turolla}}]{Sartore:2009wn}%
  \BibitemOpen
  \bibfield  {author} {\bibinfo {author} {\bibfnamefont {N.}~\bibnamefont
  {Sartore}}, \bibinfo {author} {\bibfnamefont {E.}~\bibnamefont {Ripamonti}},
  \bibinfo {author} {\bibfnamefont {A.}~\bibnamefont {Treves}}, \ and\ \bibinfo
  {author} {\bibfnamefont {R.}~\bibnamefont {Turolla}},\ }\href {\doibase
  10.1051/0004-6361/200912222} {\bibfield  {journal} {\bibinfo  {journal}
  {Astron. Astrophys.}\ }\textbf {\bibinfo {volume} {510}},\ \bibinfo {pages}
  {A23} (\bibinfo {year} {2010})},\ \Eprint {http://arxiv.org/abs/0908.3182}
  {arXiv:0908.3182 [astro-ph.GA]} \BibitemShut {NoStop}%
\bibitem [{\citenamefont {{Dewey}}\ \emph {et~al.}(1984)\citenamefont
  {{Dewey}}, \citenamefont {{Stokes}}, \citenamefont {{Segelstein}},
  \citenamefont {{Taylor}},\ and\ \citenamefont
  {{Weisberg}}}]{1984bens.work..234D}%
  \BibitemOpen
  \bibfield  {author} {\bibinfo {author} {\bibfnamefont {R.}~\bibnamefont
  {{Dewey}}}, \bibinfo {author} {\bibfnamefont {G.}~\bibnamefont {{Stokes}}},
  \bibinfo {author} {\bibfnamefont {D.}~\bibnamefont {{Segelstein}}}, \bibinfo
  {author} {\bibfnamefont {J.}~\bibnamefont {{Taylor}}}, \ and\ \bibinfo
  {author} {\bibfnamefont {J.}~\bibnamefont {{Weisberg}}},\ }in\ \href@noop {}
  {\emph {\bibinfo {booktitle} {Birth and Evolution of Neutron Stars: Issues
  Raised by Millisecond Pulsars}}},\ \bibinfo {editor} {edited by\ \bibinfo
  {editor} {\bibfnamefont {S.~P.}\ \bibnamefont {{Reynolds}}}\ and\ \bibinfo
  {editor} {\bibfnamefont {D.~R.}\ \bibnamefont {{Stinebring}}}}\ (\bibinfo
  {year} {1984})\ p.\ \bibinfo {pages} {234}\BibitemShut {NoStop}%
\bibitem [{\citenamefont {Perley}\ \emph {et~al.}(2011)\citenamefont {Perley},
  \citenamefont {Chandler}, \citenamefont {Butler},\ and\ \citenamefont
  {Wrobel}}]{2011ApJ...739L...1P}%
  \BibitemOpen
  \bibfield  {author} {\bibinfo {author} {\bibfnamefont {R.~A.}\ \bibnamefont
  {Perley}}, \bibinfo {author} {\bibfnamefont {C.~J.}\ \bibnamefont
  {Chandler}}, \bibinfo {author} {\bibfnamefont {B.~J.}\ \bibnamefont
  {Butler}}, \ and\ \bibinfo {author} {\bibfnamefont {J.~M.}\ \bibnamefont
  {Wrobel}},\ }\href {\doibase 10.1088/2041-8205/739/1/L1} {\bibfield
  {journal} {\bibinfo  {journal} {Astrophys. J. Lett.}\ }\textbf {\bibinfo
  {volume} {739}},\ \bibinfo {pages} {L1} (\bibinfo {year} {2011})},\ \Eprint
  {http://arxiv.org/abs/1106.0532} {arXiv:1106.0532 [astro-ph.IM]} \BibitemShut
  {NoStop}%
\bibitem [{\citenamefont {Calore}\ \emph {et~al.}(2016)\citenamefont {Calore},
  \citenamefont {Di~Mauro}, \citenamefont {Donato}, \citenamefont {Hessels},\
  and\ \citenamefont {Weniger}}]{Calore:2015bsx}%
  \BibitemOpen
  \bibfield  {author} {\bibinfo {author} {\bibfnamefont {F.}~\bibnamefont
  {Calore}}, \bibinfo {author} {\bibfnamefont {M.}~\bibnamefont {Di~Mauro}},
  \bibinfo {author} {\bibfnamefont {F.}~\bibnamefont {Donato}}, \bibinfo
  {author} {\bibfnamefont {J.~W.~T.}\ \bibnamefont {Hessels}}, \ and\ \bibinfo
  {author} {\bibfnamefont {C.}~\bibnamefont {Weniger}},\ }\href {\doibase
  10.3847/0004-637X/827/2/143} {\bibfield  {journal} {\bibinfo  {journal}
  {Astrophys. J.}\ }\textbf {\bibinfo {volume} {827}},\ \bibinfo {pages} {143}
  (\bibinfo {year} {2016})},\ \Eprint {http://arxiv.org/abs/1512.06825}
  {arXiv:1512.06825 [astro-ph.HE]} \BibitemShut {NoStop}%
\bibitem [{\citenamefont {Gaggero}\ \emph {et~al.}(2017)\citenamefont
  {Gaggero}, \citenamefont {Bertone}, \citenamefont {Calore}, \citenamefont
  {Connors}, \citenamefont {Lovell}, \citenamefont {Markoff},\ and\
  \citenamefont {Storm}}]{Gaggero:2016dpq}%
  \BibitemOpen
  \bibfield  {author} {\bibinfo {author} {\bibfnamefont {D.}~\bibnamefont
  {Gaggero}}, \bibinfo {author} {\bibfnamefont {G.}~\bibnamefont {Bertone}},
  \bibinfo {author} {\bibfnamefont {F.}~\bibnamefont {Calore}}, \bibinfo
  {author} {\bibfnamefont {R.~M.~T.}\ \bibnamefont {Connors}}, \bibinfo
  {author} {\bibfnamefont {M.}~\bibnamefont {Lovell}}, \bibinfo {author}
  {\bibfnamefont {S.}~\bibnamefont {Markoff}}, \ and\ \bibinfo {author}
  {\bibfnamefont {E.}~\bibnamefont {Storm}},\ }\href {\doibase
  10.1103/PhysRevLett.118.241101} {\bibfield  {journal} {\bibinfo  {journal}
  {Phys. Rev. Lett.}\ }\textbf {\bibinfo {volume} {118}},\ \bibinfo {pages}
  {241101} (\bibinfo {year} {2017})},\ \Eprint
  {http://arxiv.org/abs/1612.00457} {arXiv:1612.00457 [astro-ph.HE]}
  \BibitemShut {NoStop}%
\bibitem [{\citenamefont {Faucher-Giguere}\ and\ \citenamefont
  {Kaspi}(2006)}]{FaucherGiguere:2005ny}%
  \BibitemOpen
  \bibfield  {author} {\bibinfo {author} {\bibfnamefont {C.-A.}\ \bibnamefont
  {Faucher-Giguere}}\ and\ \bibinfo {author} {\bibfnamefont {V.~M.}\
  \bibnamefont {Kaspi}},\ }\href {\doibase 10.1086/501516} {\bibfield
  {journal} {\bibinfo  {journal} {Astrophys. J.}\ }\textbf {\bibinfo {volume}
  {643}},\ \bibinfo {pages} {332} (\bibinfo {year} {2006})},\ \Eprint
  {http://arxiv.org/abs/astro-ph/0512585} {arXiv:astro-ph/0512585} \BibitemShut
  {NoStop}%
\bibitem [{\citenamefont {Bates}\ \emph {et~al.}(2014)\citenamefont {Bates},
  \citenamefont {Lorimer}, \citenamefont {Rane},\ and\ \citenamefont
  {Swiggum}}]{Bates:2013uma}%
  \BibitemOpen
  \bibfield  {author} {\bibinfo {author} {\bibfnamefont {S.}~\bibnamefont
  {Bates}}, \bibinfo {author} {\bibfnamefont {D.}~\bibnamefont {Lorimer}},
  \bibinfo {author} {\bibfnamefont {A.}~\bibnamefont {Rane}}, \ and\ \bibinfo
  {author} {\bibfnamefont {J.}~\bibnamefont {Swiggum}},\ }\href {\doibase
  10.1093/mnras/stu157} {\bibfield  {journal} {\bibinfo  {journal} {Mon. Not.
  Roy. Astron. Soc.}\ }\textbf {\bibinfo {volume} {439}},\ \bibinfo {pages}
  {2893} (\bibinfo {year} {2014})},\ \Eprint {http://arxiv.org/abs/1311.3427}
  {arXiv:1311.3427 [astro-ph.IM]} \BibitemShut {NoStop}%
\bibitem [{\citenamefont {Lorimer}\ \emph {et~al.}(2006)\citenamefont {Lorimer}
  \emph {et~al.}}]{Lorimer:2006qs}%
  \BibitemOpen
  \bibfield  {author} {\bibinfo {author} {\bibfnamefont {D.~R.}\ \bibnamefont
  {Lorimer}} \emph {et~al.},\ }\href {\doibase
  10.1111/j.1365-2966.2006.10887.x} {\bibfield  {journal} {\bibinfo  {journal}
  {Mon. Not. Roy. Astron. Soc.}\ }\textbf {\bibinfo {volume} {372}},\ \bibinfo
  {pages} {777} (\bibinfo {year} {2006})},\ \Eprint
  {http://arxiv.org/abs/astro-ph/0607640} {arXiv:astro-ph/0607640} \BibitemShut
  {NoStop}%
\bibitem [{\citenamefont {{SKA Science Working Group}}(2012)}]{SKA-design}%
  \BibitemOpen
  \bibfield  {author} {\bibinfo {author} {\bibnamefont {{SKA Science Working
  Group}}},\ }\href@noop {} {\enquote {\bibinfo {title}
  {\href{https://www.skatelescope.org/wp-content/uploads/2013/03/SCI-020.010.020-DRM-002-3aFINAL.pdf}{The
  Square Kilometre Array Design Reference Mission: SKA Phase 1}},}\ } (\bibinfo
  {year} {2012})\BibitemShut {NoStop}%
\bibitem [{\citenamefont {{Braun}}\ \emph {et~al.}(2019)\citenamefont
  {{Braun}}, \citenamefont {{Bonaldi}}, \citenamefont {{Bourke}}, \citenamefont
  {{Keane}},\ and\ \citenamefont {{Wagg}}}]{2019arXiv191212699B}%
  \BibitemOpen
  \bibfield  {author} {\bibinfo {author} {\bibfnamefont {R.}~\bibnamefont
  {{Braun}}}, \bibinfo {author} {\bibfnamefont {A.}~\bibnamefont {{Bonaldi}}},
  \bibinfo {author} {\bibfnamefont {T.}~\bibnamefont {{Bourke}}}, \bibinfo
  {author} {\bibfnamefont {E.}~\bibnamefont {{Keane}}}, \ and\ \bibinfo
  {author} {\bibfnamefont {J.}~\bibnamefont {{Wagg}}},\ }\href@noop {}
  {\bibfield  {journal} {\bibinfo  {journal} {arXiv e-prints}\ ,\ \bibinfo
  {eid} {arXiv:1912.12699}} (\bibinfo {year} {2019})},\ \Eprint
  {http://arxiv.org/abs/1912.12699} {arXiv:1912.12699 [astro-ph.IM]}
  \BibitemShut {NoStop}%
\bibitem [{\citenamefont {Leroy}\ \emph {et~al.}(2020)\citenamefont {Leroy},
  \citenamefont {Chianese}, \citenamefont {Edwards},\ and\ \citenamefont
  {Weniger}}]{Leroy:2019ghm}%
  \BibitemOpen
  \bibfield  {author} {\bibinfo {author} {\bibfnamefont {M.}~\bibnamefont
  {Leroy}}, \bibinfo {author} {\bibfnamefont {M.}~\bibnamefont {Chianese}},
  \bibinfo {author} {\bibfnamefont {T.~D.~P.}\ \bibnamefont {Edwards}}, \ and\
  \bibinfo {author} {\bibfnamefont {C.}~\bibnamefont {Weniger}},\ }\href
  {\doibase 10.1103/PhysRevD.101.123003} {\bibfield  {journal} {\bibinfo
  {journal} {Phys. Rev. D}\ }\textbf {\bibinfo {volume} {101}},\ \bibinfo
  {pages} {123003} (\bibinfo {year} {2020})},\ \Eprint
  {http://arxiv.org/abs/1912.08815} {arXiv:1912.08815 [hep-ph]} \BibitemShut
  {NoStop}%
\bibitem [{\citenamefont {{Napier}}\ \emph {et~al.}(1983)\citenamefont
  {{Napier}}, \citenamefont {{Thompson}},\ and\ \citenamefont
  {{Ekers}}}]{1983IEEEP..71.1295N}%
  \BibitemOpen
  \bibfield  {author} {\bibinfo {author} {\bibfnamefont {P.~J.}\ \bibnamefont
  {{Napier}}}, \bibinfo {author} {\bibfnamefont {A.~R.}\ \bibnamefont
  {{Thompson}}}, \ and\ \bibinfo {author} {\bibfnamefont {R.~D.}\ \bibnamefont
  {{Ekers}}},\ }\href@noop {} {\bibfield  {journal} {\bibinfo  {journal} {IEEE
  Proceedings}\ }\textbf {\bibinfo {volume} {71}},\ \bibinfo {pages} {1295}
  (\bibinfo {year} {1983})}\BibitemShut {NoStop}%
\bibitem [{\citenamefont {Darling}(2020{\natexlab{b}})}]{Darling:2020plz}%
  \BibitemOpen
  \bibfield  {author} {\bibinfo {author} {\bibfnamefont {J.}~\bibnamefont
  {Darling}},\ }\href {\doibase 10.1103/PhysRevLett.125.121103} {\bibfield
  {journal} {\bibinfo  {journal} {Phys. Rev. Lett.}\ }\textbf {\bibinfo
  {volume} {125}},\ \bibinfo {pages} {121103} (\bibinfo {year}
  {2020}{\natexlab{b}})},\ \Eprint {http://arxiv.org/abs/2008.01877}
  {arXiv:2008.01877 [astro-ph.CO]} \BibitemShut {NoStop}%
\bibitem [{\citenamefont {Chiti}\ \emph {et~al.}(2016)\citenamefont {Chiti},
  \citenamefont {Chatterjee}, \citenamefont {Wharton}, \citenamefont {Cordes},
  \citenamefont {Lazio}, \citenamefont {Kaplan}, \citenamefont {Bower},\ and\
  \citenamefont {Croft}}]{Chiti:2016xae}%
  \BibitemOpen
  \bibfield  {author} {\bibinfo {author} {\bibfnamefont {A.}~\bibnamefont
  {Chiti}}, \bibinfo {author} {\bibfnamefont {S.}~\bibnamefont {Chatterjee}},
  \bibinfo {author} {\bibfnamefont {R.}~\bibnamefont {Wharton}}, \bibinfo
  {author} {\bibfnamefont {J.}~\bibnamefont {Cordes}}, \bibinfo {author}
  {\bibfnamefont {T.~J.~W.}\ \bibnamefont {Lazio}}, \bibinfo {author}
  {\bibfnamefont {D.~L.}\ \bibnamefont {Kaplan}}, \bibinfo {author}
  {\bibfnamefont {G.~C.}\ \bibnamefont {Bower}}, \ and\ \bibinfo {author}
  {\bibfnamefont {S.}~\bibnamefont {Croft}},\ }\href {\doibase
  10.3847/0004-637X/833/1/11} {\bibfield  {journal} {\bibinfo  {journal}
  {Astrophys. J.}\ }\textbf {\bibinfo {volume} {833}},\ \bibinfo {pages} {11}
  (\bibinfo {year} {2016})},\ \Eprint {http://arxiv.org/abs/1610.00403}
  {arXiv:1610.00403 [astro-ph.HE]} \BibitemShut {NoStop}%
\bibitem [{\citenamefont {Zhao}\ \emph {et~al.}(2020)\citenamefont {Zhao},
  \citenamefont {Morris},\ and\ \citenamefont {Goss}}]{Zhao:2020evb}%
  \BibitemOpen
  \bibfield  {author} {\bibinfo {author} {\bibfnamefont {J.-H.}\ \bibnamefont
  {Zhao}}, \bibinfo {author} {\bibfnamefont {M.~R.}\ \bibnamefont {Morris}}, \
  and\ \bibinfo {author} {\bibfnamefont {W.~M.}\ \bibnamefont {Goss}},\ }\href
  {\doibase 10.3847/1538-4357/abc75e} {\bibfield  {journal} {\bibinfo
  {journal} {Astrophys. J.}\ }\textbf {\bibinfo {volume} {905}},\ \bibinfo
  {pages} {173} (\bibinfo {year} {2020})},\ \Eprint
  {http://arxiv.org/abs/2011.01368} {arXiv:2011.01368 [astro-ph.HE]}
  \BibitemShut {NoStop}%
\bibitem [{\citenamefont {Xiao}\ \emph {et~al.}(2021)\citenamefont {Xiao},
  \citenamefont {Williams},\ and\ \citenamefont {McQuinn}}]{Xiao:2021nkb}%
  \BibitemOpen
  \bibfield  {author} {\bibinfo {author} {\bibfnamefont {H.}~\bibnamefont
  {Xiao}}, \bibinfo {author} {\bibfnamefont {I.}~\bibnamefont {Williams}}, \
  and\ \bibinfo {author} {\bibfnamefont {M.}~\bibnamefont {McQuinn}},\
  }\href@noop {} {\  (\bibinfo {year} {2021})},\ \Eprint
  {http://arxiv.org/abs/2101.04177} {arXiv:2101.04177 [astro-ph.CO]}
  \BibitemShut {NoStop}%
\bibitem [{\citenamefont {Battye}\ \emph {et~al.}(2020)\citenamefont {Battye},
  \citenamefont {Garbrecht}, \citenamefont {McDonald}, \citenamefont {Pace},\
  and\ \citenamefont {Srinivasan}}]{Battye:2019aco}%
  \BibitemOpen
  \bibfield  {author} {\bibinfo {author} {\bibfnamefont {R.~A.}\ \bibnamefont
  {Battye}}, \bibinfo {author} {\bibfnamefont {B.}~\bibnamefont {Garbrecht}},
  \bibinfo {author} {\bibfnamefont {J.~I.}\ \bibnamefont {McDonald}}, \bibinfo
  {author} {\bibfnamefont {F.}~\bibnamefont {Pace}}, \ and\ \bibinfo {author}
  {\bibfnamefont {S.}~\bibnamefont {Srinivasan}},\ }\href {\doibase
  10.1103/PhysRevD.102.023504} {\bibfield  {journal} {\bibinfo  {journal}
  {Phys. Rev. D}\ }\textbf {\bibinfo {volume} {102}},\ \bibinfo {pages}
  {023504} (\bibinfo {year} {2020})},\ \Eprint
  {http://arxiv.org/abs/1910.11907} {arXiv:1910.11907 [astro-ph.CO]}
  \BibitemShut {NoStop}%
\bibitem [{\citenamefont {Oliphant}(06  )}]{numpy}%
  \BibitemOpen
  \bibfield  {author} {\bibinfo {author} {\bibfnamefont {T.}~\bibnamefont
  {Oliphant}},\ }\href {http://www.numpy.org/} {\enquote {\bibinfo {title}
  {{NumPy}: A guide to {NumPy}},}\ }\bibinfo {howpublished} {USA: Trelgol
  Publishing} (\bibinfo {year} {2006--}),\ \bibinfo {note} {[Online; accessed
  <today>]}\BibitemShut {NoStop}%
\bibitem [{\citenamefont {{Virtanen}}\ \emph {et~al.}(2020)\citenamefont
  {{Virtanen}}, \citenamefont {{Gommers}}, \citenamefont {{Oliphant}},
  \citenamefont {{Haberland}}, \citenamefont {{Reddy}}, \citenamefont
  {{Cournapeau}}, \citenamefont {{Burovski}}, \citenamefont {{Peterson}},
  \citenamefont {{Weckesser}}, \citenamefont {{Bright}}, \citenamefont {{van
  der Walt}}, \citenamefont {{Brett}}, \citenamefont {{Wilson}}, \citenamefont
  {{Jarrod Millman}}, \citenamefont {{Mayorov}}, \citenamefont {{Nelson}},
  \citenamefont {{Jones}}, \citenamefont {{Kern}}, \citenamefont {{Larson}},
  \citenamefont {{Carey}}, \citenamefont {{Polat}}, \citenamefont {{Feng}},
  \citenamefont {{Moore}}, \citenamefont {{Vand erPlas}}, \citenamefont
  {{Laxalde}}, \citenamefont {{Perktold}}, \citenamefont {{Cimrman}},
  \citenamefont {{Henriksen}}, \citenamefont {{Quintero}}, \citenamefont
  {{Harris}}, \citenamefont {{Archibald}}, \citenamefont {{Ribeiro}},
  \citenamefont {{Pedregosa}}, \citenamefont {{van Mulbregt}},\ and\
  \citenamefont {{Contributors}}}]{scipy}%
  \BibitemOpen
  \bibfield  {author} {\bibinfo {author} {\bibfnamefont {P.}~\bibnamefont
  {{Virtanen}}}, \bibinfo {author} {\bibfnamefont {R.}~\bibnamefont
  {{Gommers}}}, \bibinfo {author} {\bibfnamefont {T.~E.}\ \bibnamefont
  {{Oliphant}}}, \bibinfo {author} {\bibfnamefont {M.}~\bibnamefont
  {{Haberland}}}, \bibinfo {author} {\bibfnamefont {T.}~\bibnamefont
  {{Reddy}}}, \bibinfo {author} {\bibfnamefont {D.}~\bibnamefont
  {{Cournapeau}}}, \bibinfo {author} {\bibfnamefont {E.}~\bibnamefont
  {{Burovski}}}, \bibinfo {author} {\bibfnamefont {P.}~\bibnamefont
  {{Peterson}}}, \bibinfo {author} {\bibfnamefont {W.}~\bibnamefont
  {{Weckesser}}}, \bibinfo {author} {\bibfnamefont {J.}~\bibnamefont
  {{Bright}}}, \bibinfo {author} {\bibfnamefont {S.~J.}\ \bibnamefont {{van der
  Walt}}}, \bibinfo {author} {\bibfnamefont {M.}~\bibnamefont {{Brett}}},
  \bibinfo {author} {\bibfnamefont {J.}~\bibnamefont {{Wilson}}}, \bibinfo
  {author} {\bibfnamefont {K.}~\bibnamefont {{Jarrod Millman}}}, \bibinfo
  {author} {\bibfnamefont {N.}~\bibnamefont {{Mayorov}}}, \bibinfo {author}
  {\bibfnamefont {A.~R.~J.}\ \bibnamefont {{Nelson}}}, \bibinfo {author}
  {\bibfnamefont {E.}~\bibnamefont {{Jones}}}, \bibinfo {author} {\bibfnamefont
  {R.}~\bibnamefont {{Kern}}}, \bibinfo {author} {\bibfnamefont
  {E.}~\bibnamefont {{Larson}}}, \bibinfo {author} {\bibfnamefont
  {C.}~\bibnamefont {{Carey}}}, \bibinfo {author} {\bibfnamefont
  {{\.I}.}~\bibnamefont {{Polat}}}, \bibinfo {author} {\bibfnamefont
  {Y.}~\bibnamefont {{Feng}}}, \bibinfo {author} {\bibfnamefont {E.~W.}\
  \bibnamefont {{Moore}}}, \bibinfo {author} {\bibfnamefont {J.}~\bibnamefont
  {{Vand erPlas}}}, \bibinfo {author} {\bibfnamefont {D.}~\bibnamefont
  {{Laxalde}}}, \bibinfo {author} {\bibfnamefont {J.}~\bibnamefont
  {{Perktold}}}, \bibinfo {author} {\bibfnamefont {R.}~\bibnamefont
  {{Cimrman}}}, \bibinfo {author} {\bibfnamefont {I.}~\bibnamefont
  {{Henriksen}}}, \bibinfo {author} {\bibfnamefont {E.~A.}\ \bibnamefont
  {{Quintero}}}, \bibinfo {author} {\bibfnamefont {C.~R.}\ \bibnamefont
  {{Harris}}}, \bibinfo {author} {\bibfnamefont {A.~M.}\ \bibnamefont
  {{Archibald}}}, \bibinfo {author} {\bibfnamefont {A.~H.}\ \bibnamefont
  {{Ribeiro}}}, \bibinfo {author} {\bibfnamefont {F.}~\bibnamefont
  {{Pedregosa}}}, \bibinfo {author} {\bibfnamefont {P.}~\bibnamefont {{van
  Mulbregt}}}, \ and\ \bibinfo {author} {\bibnamefont {{Contributors}}},\
  }\href {\doibase https://doi.org/10.1038/s41592-019-0686-2} {\bibfield
  {journal} {\bibinfo  {journal} {Nature Methods}\ }\textbf {\bibinfo {volume}
  {17}},\ \bibinfo {pages} {261} (\bibinfo {year} {2020})}\BibitemShut
  {NoStop}%
\bibitem [{\citenamefont {Hunter}(2007)}]{Hunter:2007}%
  \BibitemOpen
  \bibfield  {author} {\bibinfo {author} {\bibfnamefont {J.~D.}\ \bibnamefont
  {Hunter}},\ }\href {\doibase 10.1109/MCSE.2007.55} {\bibfield  {journal}
  {\bibinfo  {journal} {Computing in Science \& Engineering}\ }\textbf
  {\bibinfo {volume} {9}},\ \bibinfo {pages} {90} (\bibinfo {year}
  {2007})}\BibitemShut {NoStop}%
\bibitem [{\citenamefont {Manchester}\ \emph {et~al.}(2005)\citenamefont
  {Manchester}, \citenamefont {Hobbs}, \citenamefont {Teoh},\ and\
  \citenamefont {Hobbs}}]{Manchester:2004bp}%
  \BibitemOpen
  \bibfield  {author} {\bibinfo {author} {\bibfnamefont {R.~N.}\ \bibnamefont
  {Manchester}}, \bibinfo {author} {\bibfnamefont {G.~B.}\ \bibnamefont
  {Hobbs}}, \bibinfo {author} {\bibfnamefont {A.}~\bibnamefont {Teoh}}, \ and\
  \bibinfo {author} {\bibfnamefont {M.}~\bibnamefont {Hobbs}},\ }\href
  {\doibase 10.1086/428488} {\bibfield  {journal} {\bibinfo  {journal} {Astron.
  J.}\ }\textbf {\bibinfo {volume} {129}},\ \bibinfo {pages} {1993} (\bibinfo
  {year} {2005})},\ \Eprint {http://arxiv.org/abs/astro-ph/0412641}
  {arXiv:astro-ph/0412641} \BibitemShut {NoStop}%
\bibitem [{\citenamefont {Binney}\ \emph {et~al.}(1997)\citenamefont {Binney},
  \citenamefont {Gerhard},\ and\ \citenamefont {Spergel}}]{Binney:1996sv}%
  \BibitemOpen
  \bibfield  {author} {\bibinfo {author} {\bibfnamefont {J.}~\bibnamefont
  {Binney}}, \bibinfo {author} {\bibfnamefont {O.}~\bibnamefont {Gerhard}}, \
  and\ \bibinfo {author} {\bibfnamefont {D.}~\bibnamefont {Spergel}},\ }\href
  {\doibase 10.1093/mnras/288.2.365} {\bibfield  {journal} {\bibinfo  {journal}
  {Mon. Not. Roy. Astron. Soc.}\ }\textbf {\bibinfo {volume} {288}},\ \bibinfo
  {pages} {365} (\bibinfo {year} {1997})},\ \Eprint
  {http://arxiv.org/abs/astro-ph/9609066} {arXiv:astro-ph/9609066} \BibitemShut
  {NoStop}%
\bibitem [{\citenamefont {Bissantz}\ and\ \citenamefont
  {Gerhard}(2002)}]{Bissantz:2001wx}%
  \BibitemOpen
  \bibfield  {author} {\bibinfo {author} {\bibfnamefont {N.}~\bibnamefont
  {Bissantz}}\ and\ \bibinfo {author} {\bibfnamefont {O.}~\bibnamefont
  {Gerhard}},\ }\href {\doibase 10.1046/j.1365-8711.2002.05116.x} {\bibfield
  {journal} {\bibinfo  {journal} {Mon. Not. Roy. Astron. Soc.}\ }\textbf
  {\bibinfo {volume} {330}},\ \bibinfo {pages} {591} (\bibinfo {year}
  {2002})},\ \Eprint {http://arxiv.org/abs/astro-ph/0110368}
  {arXiv:astro-ph/0110368} \BibitemShut {NoStop}%
\bibitem [{\citenamefont {McMillan}(2011)}]{McMillan:2011wd}%
  \BibitemOpen
  \bibfield  {author} {\bibinfo {author} {\bibfnamefont {P.~J.}\ \bibnamefont
  {McMillan}},\ }\href {\doibase 10.1111/j.1365-2966.2011.18564.x} {\bibfield
  {journal} {\bibinfo  {journal} {Mon. Not. Roy. Astron. Soc.}\ }\textbf
  {\bibinfo {volume} {414}},\ \bibinfo {pages} {2446} (\bibinfo {year}
  {2011})},\ \Eprint {http://arxiv.org/abs/1102.4340} {arXiv:1102.4340
  [astro-ph.GA]} \BibitemShut {NoStop}%
\bibitem [{\citenamefont {Bartels}\ \emph {et~al.}(2018)\citenamefont
  {Bartels}, \citenamefont {Edwards},\ and\ \citenamefont
  {Weniger}}]{Bartels:2018xom}%
  \BibitemOpen
  \bibfield  {author} {\bibinfo {author} {\bibfnamefont {R.~T.}\ \bibnamefont
  {Bartels}}, \bibinfo {author} {\bibfnamefont {T.~D.~P.}\ \bibnamefont
  {Edwards}}, \ and\ \bibinfo {author} {\bibfnamefont {C.}~\bibnamefont
  {Weniger}},\ }\href {\doibase 10.1093/mnras/sty2529} {\bibfield  {journal}
  {\bibinfo  {journal} {Mon. Not. Roy. Astron. Soc.}\ }\textbf {\bibinfo
  {volume} {481}},\ \bibinfo {pages} {3966} (\bibinfo {year} {2018})},\ \Eprint
  {http://arxiv.org/abs/1805.11097} {arXiv:1805.11097 [astro-ph.HE]}
  \BibitemShut {NoStop}%
\bibitem [{\citenamefont {{Haensel}}\ \emph {et~al.}(1990)\citenamefont
  {{Haensel}}, \citenamefont {{Urpin}},\ and\ \citenamefont
  {{Iakovlev}}}]{1990A&A...229..133H}%
  \BibitemOpen
  \bibfield  {author} {\bibinfo {author} {\bibfnamefont {P.}~\bibnamefont
  {{Haensel}}}, \bibinfo {author} {\bibfnamefont {V.~A.}\ \bibnamefont
  {{Urpin}}}, \ and\ \bibinfo {author} {\bibfnamefont {D.~G.}\ \bibnamefont
  {{Iakovlev}}},\ }\href@noop {} {\bibfield  {journal} {\bibinfo  {journal}
  {{Astron. Astrophys.}}\ }\textbf {\bibinfo {volume} {229}},\ \bibinfo {pages}
  {133} (\bibinfo {year} {1990})}\BibitemShut {NoStop}%
\bibitem [{\citenamefont {{Goldreich}}\ and\ \citenamefont
  {{Reisenegger}}(1992)}]{1992ApJ...395..250G}%
  \BibitemOpen
  \bibfield  {author} {\bibinfo {author} {\bibfnamefont {P.}~\bibnamefont
  {{Goldreich}}}\ and\ \bibinfo {author} {\bibfnamefont {A.}~\bibnamefont
  {{Reisenegger}}},\ }\href {\doibase 10.1086/171646} {\bibfield  {journal}
  {\bibinfo  {journal} {Astrophys. J.}\ }\textbf {\bibinfo {volume} {395}},\
  \bibinfo {pages} {250} (\bibinfo {year} {1992})}\BibitemShut {NoStop}%
\bibitem [{\citenamefont {{Shalybkov}}\ and\ \citenamefont
  {{Urpin}}(1995)}]{1995MNRAS.273..643S}%
  \BibitemOpen
  \bibfield  {author} {\bibinfo {author} {\bibfnamefont {D.~A.}\ \bibnamefont
  {{Shalybkov}}}\ and\ \bibinfo {author} {\bibfnamefont {V.~A.}\ \bibnamefont
  {{Urpin}}},\ }\href {\doibase 10.1093/mnras/273.3.643} {\bibfield  {journal}
  {\bibinfo  {journal} {{Mon. Not. Roy. Astron. Soc.}}\ }\textbf {\bibinfo
  {volume} {273}},\ \bibinfo {pages} {643} (\bibinfo {year}
  {1995})}\BibitemShut {NoStop}%
\bibitem [{\citenamefont {Pons}\ and\ \citenamefont
  {Geppert}(2007)}]{Pons:2007vf}%
  \BibitemOpen
  \bibfield  {author} {\bibinfo {author} {\bibfnamefont {J.~A.}\ \bibnamefont
  {Pons}}\ and\ \bibinfo {author} {\bibfnamefont {U.}~\bibnamefont {Geppert}},\
  }\href {\doibase 10.1051/0004-6361:20077456} {\bibfield  {journal} {\bibinfo
  {journal} {Astron. Astrophys.}\ }\textbf {\bibinfo {volume} {470}},\ \bibinfo
  {pages} {303} (\bibinfo {year} {2007})},\ \Eprint
  {http://arxiv.org/abs/astro-ph/0703267} {arXiv:astro-ph/0703267} \BibitemShut
  {NoStop}%
\bibitem [{\citenamefont {O'Hare}\ and\ \citenamefont
  {Green}(2017)}]{OHare:2017yze}%
  \BibitemOpen
  \bibfield  {author} {\bibinfo {author} {\bibfnamefont {C.~A.~J.}\
  \bibnamefont {O'Hare}}\ and\ \bibinfo {author} {\bibfnamefont {A.~M.}\
  \bibnamefont {Green}},\ }\href {\doibase 10.1103/PhysRevD.95.063017}
  {\bibfield  {journal} {\bibinfo  {journal} {Phys. Rev. D}\ }\textbf {\bibinfo
  {volume} {95}},\ \bibinfo {pages} {063017} (\bibinfo {year} {2017})},\
  \Eprint {http://arxiv.org/abs/1701.03118} {arXiv:1701.03118 [astro-ph.CO]}
  \BibitemShut {NoStop}%
\bibitem [{\citenamefont {Tkachev}(1991)}]{Tkachev:1991ka}%
  \BibitemOpen
  \bibfield  {author} {\bibinfo {author} {\bibfnamefont {I.~I.}\ \bibnamefont
  {Tkachev}},\ }\href {\doibase 10.1016/0370-2693(91)90330-S} {\bibfield
  {journal} {\bibinfo  {journal} {Phys. Lett. B}\ }\textbf {\bibinfo {volume}
  {261}},\ \bibinfo {pages} {289} (\bibinfo {year} {1991})}\BibitemShut
  {NoStop}%
\bibitem [{\citenamefont {Kaup}(1968)}]{Kaup:1968zz}%
  \BibitemOpen
  \bibfield  {author} {\bibinfo {author} {\bibfnamefont {D.~J.}\ \bibnamefont
  {Kaup}},\ }\href {\doibase 10.1103/PhysRev.172.1331} {\bibfield  {journal}
  {\bibinfo  {journal} {Phys. Rev.}\ }\textbf {\bibinfo {volume} {172}},\
  \bibinfo {pages} {1331} (\bibinfo {year} {1968})}\BibitemShut {NoStop}%
\bibitem [{\citenamefont {Ruffini}\ and\ \citenamefont
  {Bonazzola}(1969)}]{Ruffini:1969qy}%
  \BibitemOpen
  \bibfield  {author} {\bibinfo {author} {\bibfnamefont {R.}~\bibnamefont
  {Ruffini}}\ and\ \bibinfo {author} {\bibfnamefont {S.}~\bibnamefont
  {Bonazzola}},\ }\href {\doibase 10.1103/PhysRev.187.1767} {\bibfield
  {journal} {\bibinfo  {journal} {Phys. Rev.}\ }\textbf {\bibinfo {volume}
  {187}},\ \bibinfo {pages} {1767} (\bibinfo {year} {1969})}\BibitemShut
  {NoStop}%
\bibitem [{\citenamefont {Colpi}\ \emph {et~al.}(1986)\citenamefont {Colpi},
  \citenamefont {Shapiro},\ and\ \citenamefont {Wasserman}}]{Colpi:1986ye}%
  \BibitemOpen
  \bibfield  {author} {\bibinfo {author} {\bibfnamefont {M.}~\bibnamefont
  {Colpi}}, \bibinfo {author} {\bibfnamefont {S.~L.}\ \bibnamefont {Shapiro}},
  \ and\ \bibinfo {author} {\bibfnamefont {I.}~\bibnamefont {Wasserman}},\
  }\href {\doibase 10.1103/PhysRevLett.57.2485} {\bibfield  {journal} {\bibinfo
   {journal} {Phys. Rev. Lett.}\ }\textbf {\bibinfo {volume} {57}},\ \bibinfo
  {pages} {2485} (\bibinfo {year} {1986})}\BibitemShut {NoStop}%
\bibitem [{\citenamefont {Raby}(2016)}]{Raby:2016deh}%
  \BibitemOpen
  \bibfield  {author} {\bibinfo {author} {\bibfnamefont {S.}~\bibnamefont
  {Raby}},\ }\href {\doibase 10.1103/PhysRevD.94.103004} {\bibfield  {journal}
  {\bibinfo  {journal} {Phys. Rev. D}\ }\textbf {\bibinfo {volume} {94}},\
  \bibinfo {pages} {103004} (\bibinfo {year} {2016})},\ \Eprint
  {http://arxiv.org/abs/1609.01694} {arXiv:1609.01694 [hep-ph]} \BibitemShut
  {NoStop}%
\bibitem [{\citenamefont {Dietrich}\ \emph {et~al.}(2019)\citenamefont
  {Dietrich}, \citenamefont {Day}, \citenamefont {Clough}, \citenamefont
  {Coughlin},\ and\ \citenamefont {Niemeyer}}]{Dietrich:2018jov}%
  \BibitemOpen
  \bibfield  {author} {\bibinfo {author} {\bibfnamefont {T.}~\bibnamefont
  {Dietrich}}, \bibinfo {author} {\bibfnamefont {F.}~\bibnamefont {Day}},
  \bibinfo {author} {\bibfnamefont {K.}~\bibnamefont {Clough}}, \bibinfo
  {author} {\bibfnamefont {M.}~\bibnamefont {Coughlin}}, \ and\ \bibinfo
  {author} {\bibfnamefont {J.}~\bibnamefont {Niemeyer}},\ }\href {\doibase
  10.1093/mnras/sty3158} {\bibfield  {journal} {\bibinfo  {journal} {Mon. Not.
  Roy. Astron. Soc.}\ }\textbf {\bibinfo {volume} {483}},\ \bibinfo {pages}
  {908} (\bibinfo {year} {2019})},\ \Eprint {http://arxiv.org/abs/1808.04746}
  {arXiv:1808.04746 [astro-ph.HE]} \BibitemShut {NoStop}%
\bibitem [{\citenamefont {Visinelli}\ \emph {et~al.}(2018)\citenamefont
  {Visinelli}, \citenamefont {Baum}, \citenamefont {Redondo}, \citenamefont
  {Freese},\ and\ \citenamefont {Wilczek}}]{Visinelli:2017ooc}%
  \BibitemOpen
  \bibfield  {author} {\bibinfo {author} {\bibfnamefont {L.}~\bibnamefont
  {Visinelli}}, \bibinfo {author} {\bibfnamefont {S.}~\bibnamefont {Baum}},
  \bibinfo {author} {\bibfnamefont {J.}~\bibnamefont {Redondo}}, \bibinfo
  {author} {\bibfnamefont {K.}~\bibnamefont {Freese}}, \ and\ \bibinfo {author}
  {\bibfnamefont {F.}~\bibnamefont {Wilczek}},\ }\href {\doibase
  10.1016/j.physletb.2017.12.010} {\bibfield  {journal} {\bibinfo  {journal}
  {Phys. Lett. B}\ }\textbf {\bibinfo {volume} {777}},\ \bibinfo {pages} {64}
  (\bibinfo {year} {2018})},\ \Eprint {http://arxiv.org/abs/1710.08910}
  {arXiv:1710.08910 [astro-ph.CO]} \BibitemShut {NoStop}%
\bibitem [{\citenamefont {Schive}\ \emph {et~al.}(2014)\citenamefont {Schive},
  \citenamefont {Liao}, \citenamefont {Woo}, \citenamefont {Wong},
  \citenamefont {Chiueh}, \citenamefont {Broadhurst},\ and\ \citenamefont
  {Hwang}}]{Schive:2014hza}%
  \BibitemOpen
  \bibfield  {author} {\bibinfo {author} {\bibfnamefont {H.-Y.}\ \bibnamefont
  {Schive}}, \bibinfo {author} {\bibfnamefont {M.-H.}\ \bibnamefont {Liao}},
  \bibinfo {author} {\bibfnamefont {T.-P.}\ \bibnamefont {Woo}}, \bibinfo
  {author} {\bibfnamefont {S.-K.}\ \bibnamefont {Wong}}, \bibinfo {author}
  {\bibfnamefont {T.}~\bibnamefont {Chiueh}}, \bibinfo {author} {\bibfnamefont
  {T.}~\bibnamefont {Broadhurst}}, \ and\ \bibinfo {author} {\bibfnamefont
  {W.~Y.~P.}\ \bibnamefont {Hwang}},\ }\href {\doibase
  10.1103/PhysRevLett.113.261302} {\bibfield  {journal} {\bibinfo  {journal}
  {Phys. Rev. Lett.}\ }\textbf {\bibinfo {volume} {113}},\ \bibinfo {pages}
  {261302} (\bibinfo {year} {2014})},\ \Eprint {http://arxiv.org/abs/1407.7762}
  {arXiv:1407.7762 [astro-ph.GA]} \BibitemShut {NoStop}%
\end{thebibliography}%

\clearpage
\newpage
\maketitle
\onecolumngrid
\begin{center}
\textbf{\large Transient Radio Signatures from Neutron Star Encounters \\ with QCD Axion Miniclusters} \\ 
\vspace{0.1in}
{ \it \large Supplemental Material}\\ 
\vspace{0.05in}
{Thomas D. P. Edwards, \ Bradley J. Kavanagh, \ Luca Visinelli, and \ Christoph Weniger}
\end{center}
\onecolumngrid
\setcounter{equation}{0}
\setcounter{figure}{0}
\setcounter{table}{0}
\setcounter{section}{0}
\setcounter{page}{1}
\makeatletter
\renewcommand{\theequation}{S\arabic{equation}}
\renewcommand{\thefigure}{S\arabic{figure}}
\subsection{Neutron Star Population}
\label{sec:NS_population}

Considerable effort has been put into modelling the population of neutron stars (NSs) in the Milky Way (MW)~\cite{FaucherGiguere:2005ny, Bates:2013uma} given the sample of those we can actually observe (see for example those reported by the Australia Telescope National Facility pulsar catalogue~\cite{Manchester:2004bp}). We assume that the spatial distribution of millisecond pulsars in the MW can be used to approximate the corresponding distribution of \textit{old} NSs, as in Ref.~\cite{Safdi:2018oeu}.

The MW hosts around $10^9$ NSs~\cite{Sartore:2009wn}, of which 20\% have been unbound due to natal kicks~\cite{Sartore:2009wn}. Of these NSs, 60\% are formed in the bulge and 40\% in the disk~\cite{Ofek:2009wt, Sartore:2009wn}. We normalize the spatial distributions in the bulge and in the disk assuming the total numbers $N_{\rm bulge} = 4.8 \times 10^8$ and $N_{\rm disk} = 3.2\times 10^8$, respectively. We model the NS spatial distributions in terms of the galactocentric cylindrical coordinates $r_\mathrm{cyl}$ and $z_\mathrm{cyl}$, which describe the radial distance from the axis of symmetry and the height from the Galactic plane respectively. Here, we assume that the spatial distribution of NSs in the bulge tracks the stellar population. We fix this in the companion paper~\cite{Kavanagh:2020gcy} as a truncated Power-law distribution~\cite{Binney:1996sv, Bissantz:2001wx}
\be
	\label{eq:NSbulgedistribution}
    n_{\rm bulge}(r_\mathrm{cyl}, z_\mathrm{cyl}) = N_{\rm bulge}\,\frac{11.1}{\rm \,kpc^3} \frac{e^{-\left(r'/r_{\rm cut}\right)^2}}{\left(1 + r'/r_0\right)^\lambda}\,,
\ee
where we use the parameters from Ref.~\cite{McMillan:2011wd}, namely the core density $\rho_0^{\rm bulge} \approx 99.3\,M_\odot/$pc$^3$, $r' = \sqrt{r_\mathrm{cyl}^2 + (z_\mathrm{cyl}/q)^2}$ with $q = 0.5$, the bulge cutoff $r_0 = 0.075{\rm \,kpc}$, the exponent $\lambda = 1.8$, and $r_{\rm cut} = 2.1\,$kpc. The numerical factor $11.1$ accounts for the integration of the NS density over the bulge volume.
Note, that this choice differs from other literature on the subject in which a Hernquist profile is assumed~\cite{Safdi:2018oeu}.

We use a Lorimer profile to model the distribution of millisecond pulsars in the Galactic disk~\cite{Lorimer:2006qs}
\be
	\label{eq:NSdiscdistribution}
	n_{\rm disk}(r_\mathrm{cyl}, z_\mathrm{cyl}) = N_{\rm disk}\,\frac{C^{B+2}\,e^{-C}}{4\pi \,r_\odot^2\sigma_z\,\Gamma(B+2)}\,\left(\frac{r_\mathrm{cyl}}{r_\odot}\right)^B\,e^{-C\frac{r_\mathrm{cyl}-r_\odot}{r_\odot}}\,e^{-\frac{|z_\mathrm{cyl}|}{\sigma_z}}\,,
\ee
with parameters that are obtained from a fit to the population of almost one hundred millisecond pulsars --- these are taken from Table III of Ref.~\cite{Bartels:2018xom}, namely $B = 3.91$, $C = 7.54$, and $\sigma_z = 0.76\,$kpc.

We have not incorporated any decay mechanisms for the NS's magnetic field, such as ohmic dissipation~\cite{1990A&A...229..133H}, ambipolar diffusion~\cite{1992ApJ...395..250G, 1995MNRAS.273..643S}, or Hall drift~\cite{Pons:2007vf}. We assume that all NSs have a mass of $M_{\rm NS} = 1.4\,M_\odot$ and radius $R_{\rm NS} = 10\,$km.

\subsection{Neutron Star Magnetosphere}
\label{sec:NS_magnetosphere}

Here, we use the Goldreich-Julian model~\cite{Goldreich:1969sb} of the NS magnetosphere, for which the magnetic field along the axis of rotation $\hat{\eta}$ is
\be
    B_{\hat{\eta}}(r,\theta_{\rm obs}) = B_0\,\left(\frac{R_{\rm NS}}{r}\right)^3\,\frac{3\cos^2\theta_{\rm obs} - 1}{2}\,,
    \label{eq:Bfield}
\ee
where the radial dependence shows the typical dipole behavior falling as $\propto r^{-3}$. For simplicity, we have assumed that the magnetic field is aligned with the axis of rotation, which are both inclined at an angle $\theta_{\rm obs} \in \left[-\pi/2,\pi/2\right]$ with respect to the observer. Each NS in the population is described by a magnetic field strength at the poles $B_0$ and a period $P$ which are drawn from log-normal distributions, with mean and dispersion given by $\log_{10}(B/{\rm G}) = 12.65$ and $\sigma_B = 0.55$ for the magnetic field strength~\cite{FaucherGiguere:2005ny, Bates:2013uma}, and $\log_{10}(P/{\rm ms}) = 2.7$ and $\sigma_P = 0.34$ for the period~\cite{Lorimer:2006qs}.

Given the angular velocity vector of the NS ${\bf \Omega}$ with absolute value $\Omega = 2\pi/P$, the charged plasma in the magnetosphere at distance $r$ has a number density~\cite{Goldreich:1969sb}
\be
    n_c = \frac{2 \Omega B_{\hat{\eta}}(r,\theta_{\rm obs})}{e} + {\rm relativistic\,\,corrections}\,.
\ee
The plasma frequency can be expressed as $\omega_p = \sqrt{4\pi\alpha_{\rm EM}n_c/m_c}$  where $\alpha_{\rm EM}$ is the fine structure constant and $m_c$ is the charge carrier mass. For electrons, we obtain
\be
    \omega_p = 150{\rm \,GHz}\,\sqrt{\left(\frac{B_{\hat{\eta}}(r,\theta_{\rm obs})}{10^{14}{\rm \,G}}\right)\left(\frac{1{\rm \,s}}{P}\right)}\,.
    \label{eq:plasma_frequency}
\ee
The conversion radius $R_c$ is defined as the region for which the plasma frequency equals the axion mass. Using Eq.~\eqref{eq:plasma_frequency}, which is valid in the electron-dominated region, the conversion radius is given by~\cite{Hook:2018iia}
\be
    R_c(\theta_\mathrm{obs}) = 224{\rm \,km}\left(\frac{R_{\rm NS}}{10{\rm \,km}}\right)\left[\left|3\cos^2\theta_\mathrm{obs}-1\right|\frac{B_0}{10^{14}{\rm \,G}}\frac{\rm 1\,s}{P}\left(\frac{\rm 1\,GHz}{m_a}\right)^2\right]^{1/3}\,,
    \label{eq:conversion_radius}
\ee
where the resonant conversion only takes place if $R_c(\theta_\mathrm{obs}) > R_{\rm NS}$. 

\subsection{Axion Minicluster Density Profiles}
\label{sec:AMCdensityprofile}

\begin{figure}[t!]
\includegraphics[width=0.49\textwidth]{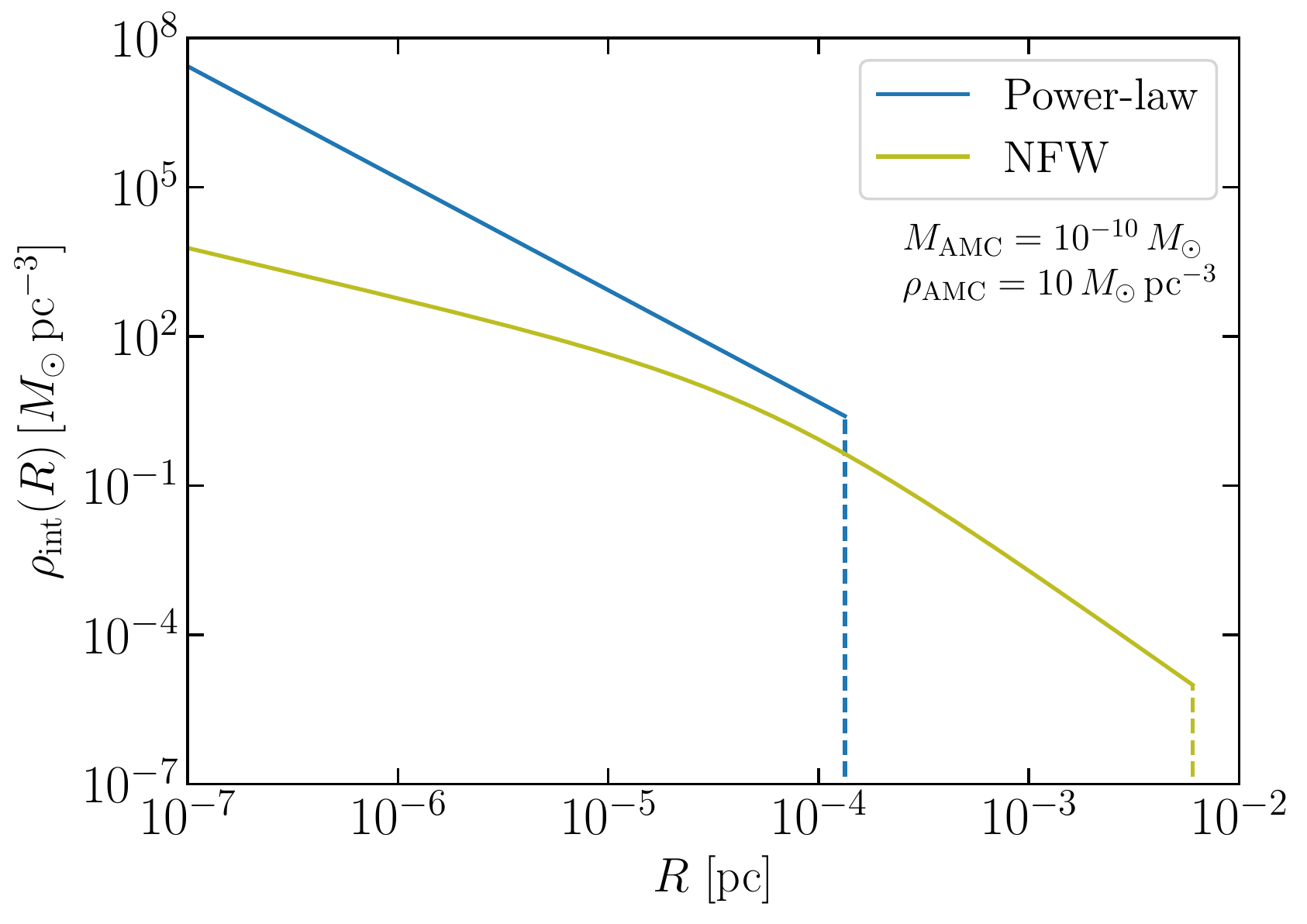}
	\caption{Models for the internal density profile's of AMCs which we consider in this paper: Power-law, Eq.~\eqref{eq:PLprofile}, and NFW, Eq.~\eqref{eq:density_NFW}. Vertical dashed lines show the truncation radii $R_\mathrm{AMC}$. We fix the characteristic mass and density to $M_{\mathrm{AMC}}=10^{-10} \, M_{\odot}$ and $\rho_{\mathrm{AMC}}=10\, M_{\odot} \,\mathrm{pc}^{-3}$ respectively.}
	\label{fig:AMC_densityprofiles}
\end{figure}

As described in main text, we use two different parameterizations for the internal density profiles of the AMCs. Since we do not know the internal density profiles precisely, these two choices are made to reflect the range of potentially observable radio signatures. An example of both density profiles and their corresponding truncation radii can be seen in Fig.~\ref{fig:AMC_densityprofiles}.

An AMC with a Power-law (PL) profile is described by~\cite{OHare:2017yze,Fairbairn:2017sil}
\be
    \label{eq:PLprofile}
    \rho_\mathrm{int}^{\mathrm{PL}}(R) = \rho_s \left(\frac{r_s}{R}\right)^{9/4}\,{\rm \Theta}\left( R_\mathrm{AMC}^{\rm PL} - R \right)\,,
\ee
 where ${\rm \Theta}\left(x\right)$ is the Heaviside step function. We truncate the PL profile at a radius 
\be
    \label{eq:RmaxPL}
    R_\mathrm{AMC}^{\rm PL} = \left(\frac{3 \MC}{4 \pi \rC(\delta)}\right)^{1/3}\,,
\ee
where we fix $\rho_s r_s^{9/4} = \rC(\delta) (R_\mathrm{AMC}^{\rm PL})^{9/4}/4$~\cite{Fairbairn:2017sil}, to give mean density $\rC(\delta)$ and the correct total mass for the AMC.

On the other hand, AMCs with NFW density profiles are described by
\be
    \rho_\mathrm{int}^{\mathrm{NFW}}(r) = \frac{\rC(\delta)}{(r/r_s)(1 + r/r_s)^2}\,, \qquad r_s = \left(\frac{\MC}{4 \pi \rC(\delta) f_{\rm NFW}(c)}\right)^{1/3}\,,
    \label{eq:density_NFW}
\ee
where the function $f_\mathrm{NFW}(c) = \ln(1 + c) - c/(1+c)$ is defined in terms of a concentration parameter $c \approx 100$~\cite{Eggemeier:2019khm,Ellis:2020gtq}. The truncation radius is now given by $R_\mathrm{AMC}^{\rm NFW} = c\, r_s$.

\subsection{Flux Distributions}
\label{sec:flux_distribution}

\begin{figure*}[h!]
	\begin{center}
	\includegraphics[width=0.49\textwidth]{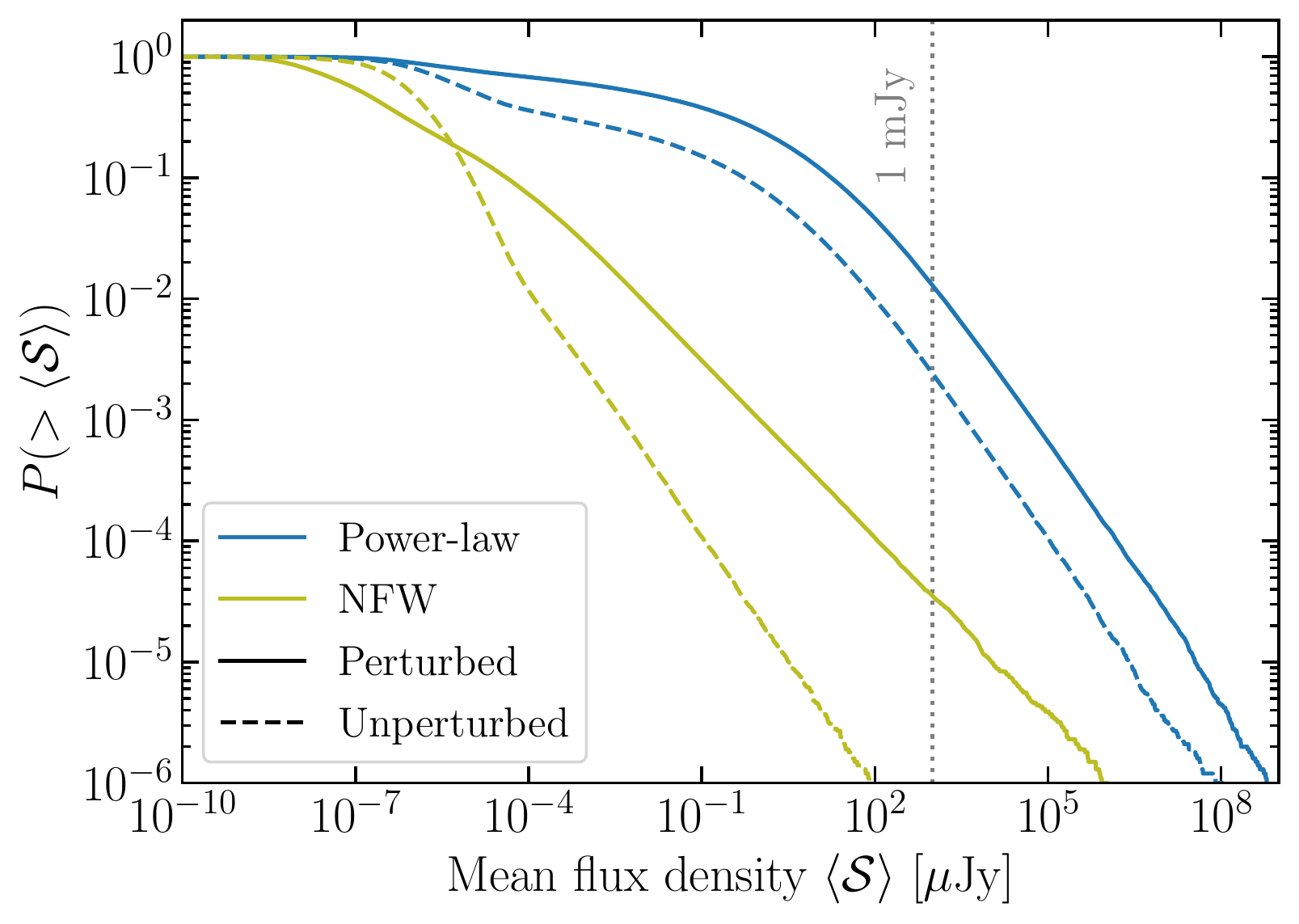}	\includegraphics[width=0.49\textwidth]{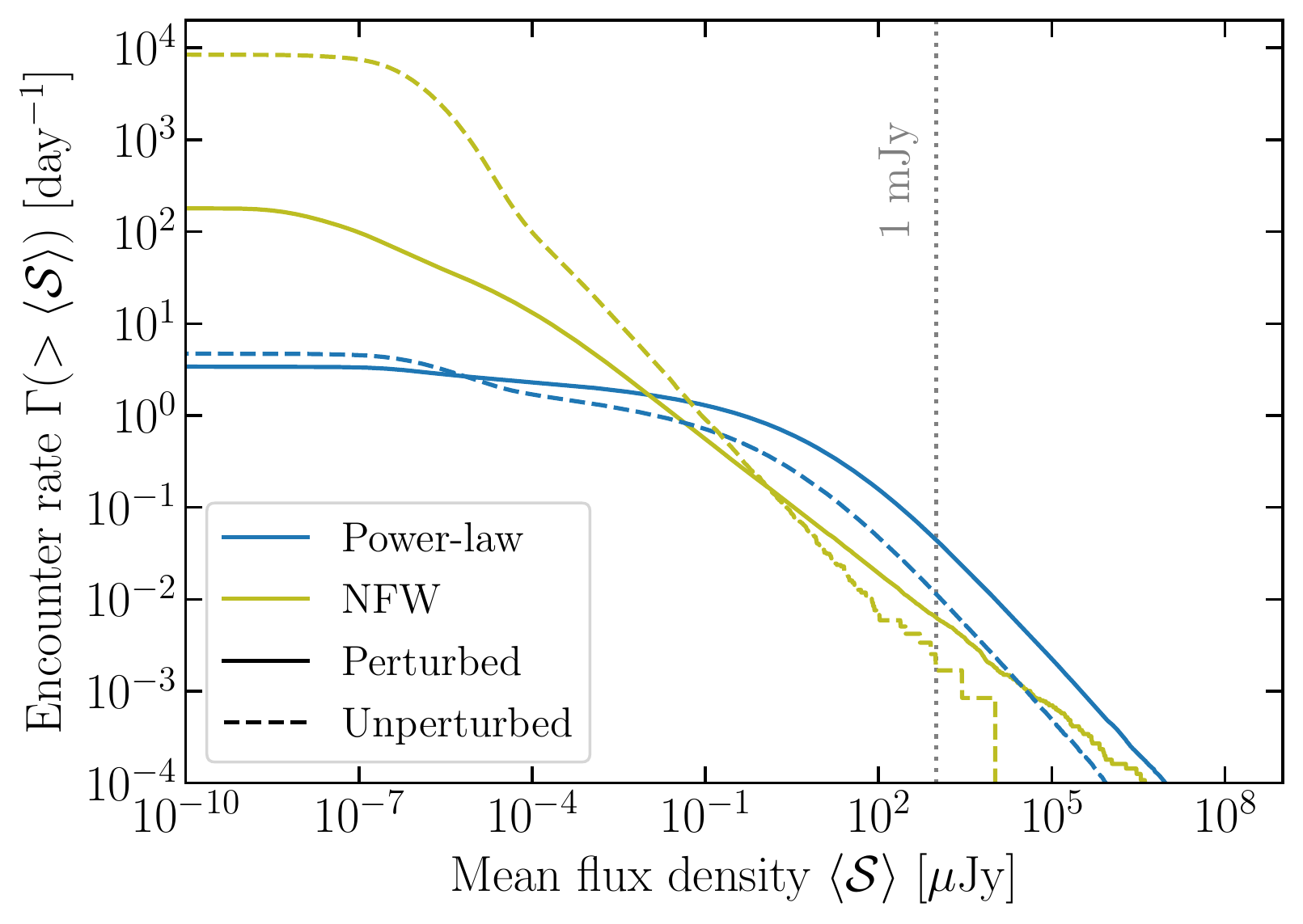}
	\caption{\textit{Left:} The cumulative probability of AMC-NS encounters as a function of the mean flux $\langle \mathcal{S}\rangle$ of the resulting radio signal. \textit{Right:} Cumulative rate of encounters above a given mean flux $\langle \mathcal{S} \rangle$.}
	\label{fig:flux}
	\end{center}
\end{figure*}

Here, we give more details concerning the expected distributions of radio fluxes from AMC-NS encounters. In the left panel of Fig.~\ref{fig:flux}, we plot the cumulative probability distribution of the mean flux density $\langle \mathcal{S} \rangle$ (that is, the fraction of events above a given value of  $\langle \mathcal{S} \rangle$). We show results for AMCs with Power-law (solid blue) and NFW (solid olive) internal density profiles. The typical flux density from an encounter between an NS and a Power-law minicluster is larger because these AMCs are substantially more dense than those with NFW profiles.  We show also the results for AMCs which have not undergone perturbations due to stellar encounters (dashed lines). For Power-law miniclusters, these results are very similar to the perturbed case; their higher density also makes them more resistant to disruption. Instead, for NFW miniclusters, the typical flux which we would expect when neglecting perturbations is much smaller than when perturbations are included. 

We can see this expressed also in terms of the encounter rate above a given threshold in flux $\langle \mathcal{S} \rangle$, as shown in the right panel of Fig.~\ref{fig:flux}. In the NFW case, going from the perturbed to unperturbed distributions, the encounter rate drops by a factor of around 40. However, the rate of very bright encounters actually \textit{increases} once perturbations are taken into account. As we show in detail in Ref.~\cite{Kavanagh:2020gcy}, the survival probability for NFW miniclusters is typically larger than 50\% throughout the MW. However, surviving AMCs are stripped of a significant fraction of their mass, typically leaving behind a much more dense remnant AMC. Thus, what would be common encounters with large, diffuse AMCs in the unperturbed case become rarer but brighter encounters with small, dense AMCs once perturbations are accounted for.

Of particular interest is that for the very brightest events (above around $1 \,\mathrm{Jy}$) the NS encounter rates for Power-law and NFW miniclusters start to converge, typically to within an order of magnitude. Despite the substantial differences in their sizes and densities, we find that the rate of bright encounters between NSs and AMCs in the MW is somewhat insensitive to the initial density profiles of the AMCs.

\subsection{The Role of Axion Stars}
\label{sec:The role of axion stars}

An axion star (AS)~\cite{Tkachev:1991ka} is a condensate made of cold axions, described by a solitonic solution of the relativistic Klein-Gordon equation~\cite{Kaup:1968zz, Ruffini:1969qy}. The axions inside the star are usually non-relativistic, so that a description in terms of the Schr{\"o}dinger-Poisson (SP) equation is often a suitable approximation. Since axions are pseudo-scalar particles, ASs differ from the analogous solutions for `boson stars' obtained in scalar boson theories~\cite{Colpi:1986ye}. While for scalar bosons a static solitonic solution to the SP exists, for pseudo-scalar axions the solution has to be oscillating periodically in time. An AS is then made up of a self-gravitating, oscillating axion field.

ASs may form in the dense central region of an axion minicluster, where the density is high enough that two-to-two processes enable the cooling of its inner core and lead to the formation of the condensate~\cite{Kolb:1993zz, Seidel:1993zk}. This process has been observed in recent numerical simulations~\cite{Levkov:2018kau,Eggemeier:2019jsu,Chen:2020cef}. The existence of ASs could be indirectly probed through their interaction with stellar  objects, which leads to a vast array of potentially detectable signals in the form of gravitational waves, neutrinos, and electromagnetic radiation~\cite{Raby:2016deh, Dietrich:2018jov}.

Equilibrium in the so-called `dilute' branch is granted by quantum pressure --- due to the wave-like nature of the axions --- which supports the AS from collapsing under its own self-gravity.
Other branches in which gravity is replaced by self-interactions have been shown to be unstable or even non-existing~\cite{Visinelli:2017ooc}. For this reason, we limit our discussion to the dilute branch, in which the radius of the AS scales inversely with the AS mass, $R_{\rm AS} \propto M_{\rm AS}^{-1}$, a relation which can be inferred from energy conservation arguments~\cite{Visinelli:2017ooc}.
The proportionality constant must be determined by numerically solving the SP equation. References~\cite{Schive:2014hza,Eggemeier:2019jsu} find
\begin{equation}
    R_{\rm AS} = 3.85\times 10^{-8}{\rm \,m} \left(\frac{20{\rm \,\mu eV}}{m_a}\right)^2\,\left(\frac{M_\odot}{M_{\rm AS}}\right)\,.
    \label{eq:axionstarradius}
\end{equation}

Simulations of DM with wave-like properties of dwarf galaxy scales suggest a relation between the mass of the solitonic core and the mass of its host halo~\cite{Schive:2014hza}. This relation has recently been confirmed for heavier axion-like particles, as we consider here, suggesting that ASs formed at the center of AMCs have a mass:
\begin{equation}
        M_{\rm AS} = 1.56 \times 10^{-13}\, M_\odot \left(\frac{20{\rm \,\mu eV}}{m_a}\right)\left(\frac{ \MC }{1 \,M_\odot}\right)^{1 / 3}\,.
        \label{eq:axionstarmass}
\end{equation}
Combining Eqs.~\eqref{eq:axionstarradius} and \eqref{eq:axionstarmass}, we can write:
\begin{equation}
    R_\mathrm{AS} =  R_\star \left( \frac{M_\mathrm{AMC}}{M_\star}\right)^{-1/3}\,,
    \label{eq:ASrelation}
\end{equation}
where for $m_a = 20 \,\mu\mathrm{eV}$ we fix the constants $R_\star = 1.7 \times 10^{-6} \,\mathrm{pc}$ and $M_\star = 10^{-16}\,M_\odot$. While current numerical simulations cannot resolve the formation of ASs in the smallest AMCs we consider, we will assume that Eq.~\eqref{eq:ASrelation} holds generally.

The presence of ASs in the centers of AMCs may affect their behaviour under stellar perturbations. We neglect this effect, which should be small for the heaviest AMCs. More dramatically, at sufficiently low AMC mass, the radius of the AS formed at the center may exceed the radius of the AMC itself. We remain agnostic about the formation and behaviour of these light AMCs and instead apply a cut (referred to as the `AS cut' in Ref.~\cite{Kavanagh:2020gcy}) which discards all AMCs for which the AS radius exceeds the AMC radius. More precisely, the AS cut therefore removes all AMCs for which:
\begin{equation}
    R_f > R_\mathrm{AS}(M_i) = R_\star \left( \frac{M_i}{M_\star}\right)^{-1/3}\,,
\end{equation}
where $M_i$ is the AMC mass before stellar perturbations are accounted for and $R_f$ is the final AMC radius after perturbations. In the main text, we present results in which we begin with $f_\mathrm{AMC} = 1$ over the full range of AMC masses $[M_\mathrm{min}, M_\mathrm{max}]$, which then undergo stellar perturbations, followed by the AS cut. As a guide, the fraction of AMCs passing the AS cut \textit{before} perturbations is $f_\mathrm{cut}^\mathrm{PL} = 2.7 \times 10^{-4}$ for PL density profiles and $f_\mathrm{cut}^\mathrm{NFW} = 1.5 \times 10^{-2}$ for NFW profiles.

\end{document}